\documentclass[twocolumn]{aastex631} 

\hypersetup{linkcolor=blue,citecolor=blue,filecolor=cyan,urlcolor=blue}

\usepackage[T1]{fontenc}
\usepackage{CJKutf8}

\begin{document}

\title{Identification of 4876 Bent-Tail Radio Galaxies in the FIRST Survey Using Deep Learning Combined with Visual Inspection}

\correspondingauthor{Baoqiang Lao}
\email{lbq19881213@gmail.com}

\author[0000-0002-3426-3269]{Baoqiang Lao}
\affiliation{School of Physics and Astronomy, Yunnan University, Kunming 650091, People's Republic of China}


\author[0000-0003-4873-1681]{Heinz Andernach}
\affiliation{Th{\"u}ringer Landessternwarte, Sternwarte 5, D-07778 Tautenburg, Germany}
\affiliation{Depto. de Astronom\'{i}a, Univ. de Guanajuato, Callej\'{o}n de Jalisco s/n, Guanajuato, C.P. 36023, GTO, Mexico}

\author[0000-0002-4439-5580]{Xiaolong Yang}
\affiliation{Shanghai Astronomical Observatory, Chinese Academy of Sciences, Shanghai 200030, People's Republic of China}

\author[0000-0002-2218-5638]{Xiang Zhang}
\affiliation{LESIA, Observatoire de Paris, Université PSL, CNRS, Sorbonne Université, Université Paris Cité, 5 place Jules Janssen, 92195 Meudon, France}

\author[0000-0002-1243-0476]{Rushuang Zhao}
\affiliation{School of Physics and Electronic Science, Guizhou Normal University, Guiyang 550001, People's Republic of China}

\author[0000-0002-0796-4078]{Zhen Zhao} 
\affiliation{Shanghai AI Laboratory, Shanghai 200003, People's Republic of China}

\author[0009-0001-3764-4307]{Yun Yu} 
\affiliation{Shanghai Astronomical Observatory, Chinese Academy of Sciences, Shanghai 200030, People's Republic of China}

\author[0000-0002-3464-5128]{Xiaohui Sun} 
\affiliation{School of Physics and Astronomy, Yunnan University, Kunming 650091, People's Republic of China}

\author[0000-0003-2302-0613]{Sheng-Li Qin} 
\affiliation{School of Physics and Astronomy, Yunnan University, Kunming 650091, People's Republic of China}







\begin{abstract}
Bent-tail radio galaxies (BTRGs) are characterized by bent radio lobes. This unique shape is mainly caused by the movement of the galaxy within a cluster, during which the radio jets are deflected by the intra-cluster medium. A combined method, which involves a deep learning-based radio source finder along with visual inspection, has been utilized to search for BTRGs from the Faint Images of the Radio Sky at Twenty-centimeters survey images. Consequently, a catalog of 4876 BTRGs has been constructed, among which 3871 are newly discovered. Based on the classification scheme of the opening angle between the two jets of the galaxy, BTRGs are typically classified as either wide-angle-tail (WAT) sources or narrow-angle-tail (NAT) sources. Our catalog comprises 4424 WATs and 652 NATs. Among these, optical counterparts are identified for 4193 BTRGs. This catalog covers luminosities in the range of $1.91\times10^{20} \leq L_{\rm 1.4\,GHz} \leq 1.45\times10^{28}$ ${\rm W\,Hz^{-1}}$ and redshifts from $z = 0.0023$ to $z = 3.43$. Various physical properties of these BTRGs and their statistics are presented. Particularly, by the nearest neighbor method, we found that 1825 BTRGs in this catalog belong to galaxy clusters reported in literature.

\end{abstract}




\section{Introduction} \label{sec:intro}
Bent-tail radio galaxies (BTRGs) \citep[e.g.][]{1968MNRAS.138....1R,1976ApJ...205L...1O} are a subclass of radio galaxy that showcase a distinctive morphology primarily influenced by their radio jets. The jet tails of the BTRG are bent away from the straight 180$^\circ$ alignment on a radio map and typically make the BTRG a `C', `V', or `U' shaped source. 
Radio galaxies with jets which are not bent in the common direction are not included in this study, such as S-shaped and Z-shaped sources, as they are bent due to different effects caused by jet precession \citep{2023ApJ...948...25N}. IC 310 and NGC 1265 were the ﬁrst two BTRGs discovered by \citet{1968MNRAS.138....1R}. 
The prototypical BTRG IC 310 \citep[e.g.][]{1998A&A...331..475F} was later discovered to be a hard gamma-ray source and was believed to be a one-sided BL Lac-type object for some time \citep{2012A&A...538L...1K} until it was reestablished as a BTRG with improved resolution observation \citep{2020MNRAS.499.5791G}.
Subsequently, the subclass of BTRG was first coined by 
\citet{1976ApJ...203L.107R}, who identified six BTRGs within rich clusters of galaxies, as observed by the National Radio Astronomy Observatory (NRAO) 3-element interferometer.

BTRGs are commonly divided into wide-angle-tail (WAT) \citep[e.g.][]{1976ApJ...205L...1O,1986ApJ...307...73B} and narrow-angle-tail (NAT) or head-tail (HT) radio galaxies \citep[e.g.][]{1976ApJ...203L.107R,1985ApJ...295...80O}, based on the opening angle ($OA$) between the two opposing jets that originate from the nucleus of the optical galaxy, where the supermassive black hole is presumed to be actively ejecting these jets. A source with an $OA$ of 180$^\circ$ implies a `straight-line' morphology, while a source with an $OA$ of 0$^\circ$ means that both jets of the source are pointing in the same direction and the source appears one-sided or single-jet. WATs have a larger $OA$ and typically have a `C-shaped' morphology, and NATs have a smaller $OA$ and generally have a `V-/U-shaped' morphology \citep{1977AJ.....82....1R}. Nonetheless, the differentiation between WATs and NATs is heavily influenced by the angular resolution and/or the distance to the radio source. For instance, the first HT source NGC 1265 \citep{1968MNRAS.138....1R} can be considered a WAT when its core region is observed at sufficiently high resolution, as seen in the left panel of Figure 1 of \citet{2021Galax...9...85R}, while the right panel shows its NAT morphology on larger scales. Extended radio galaxies are traditionally classified into Fanaroff-Riley (FR) Type I (FR-I) and FR Type II (FR-II), where sources with an ``edge-darkened'' feature which were mostly found to have a radio luminosity 
$<10^{25}\,{\rm W}\,{\rm Hz}^{-1}$ at 1.4 GHz are defined as FR-Is, while those dominated by ``edge-brightened'' features which were mostly at radio luminosities 
$>10^{25}\,{\rm W}\,{\rm Hz}^{-1}$ at 1.4 GHz are called FR-IIs \citep{1974MNRAS.167P..31F,1984ARA&A..22..319B}. However, some later studies have found that the classical luminosity break is not a distinct diagnostic characteristic \citep[e.g.][]{2017A&A...601A..81C,2019MNRAS.488.2701M}. They discovered that radio luminosities of FR-IIs existed below it, indicating that FR morphology is generally more significant than the luminosity break. NATs generally are consistent with the FR-I definition \citep{2011ApJS..194...31P}, and the WATs can be detected in either FR-Is or FR-IIs \citep{2019AA...626A...8M}. However, not all BTRGs can satisfy the definition of FR classification. Thus, treating BTRGs as a third and independent FR class is also beneficial for research on the morphological classification of radio galaxies \citep[e.g.][]{2017ApJS..230...20A,2023DIB....4708974G}.

The bent morphology of BTRGs is a direct result of the interaction between the relativistic jets emitted by the active galactic nuclei (AGN) and the surrounding interstellar medium (ISM). Studying these jets helps astrophysicists understand the mechanisms of jet launching, collimation, and propagation, as well as the effects of relativistic motion on the observed properties of these jets \citep{2014MNRAS.442..838M}. 
These structures within BTRGs can provide valuable insights into the magnetic fields both within and surrounding these galaxies \citep{2013MNRAS.432..243P}.
Additionally, BTRGs are generally found in dense environments, such as galaxy clusters \citep{1976Natur.259..451V,1976ApJ...203L.107R,1996MNRAS.282..137J,2009A&A...505...55G,2023ApJ...957L...4L}. Their presence and characteristics can shed light on the effects of environmental factors such as ram pressure, tidal forces, and interactions with the intra-cluster medium (ICM) on galaxy evolution \citep{2008ApJ...685..858F,2020MNRAS.496.4654C}. 

Over the 50 years since the initial discovery of BTRG, BTRG research has been consistently enriched and expanded upon. The BTRGs discovered during these 50 years were selected from individual observations that enabled a significant number of discoveries due to their high resolution and high sensitivity. Examples of such observations include those at a frequency of 1.4 GHz using the Very Large Array (VLA) \citep{1990ApJS...72...75O}, the VLA Faint Images of the Radio Sky at Twenty-centimeters (FIRST) survey \citep{2000PhDT........13B,2011AJ....141...88W,2011ApJS..194...31P,2017ApJ...844...78P}, and the 1.4 GHz Australia Telescope Large Area Survey (ATLAS) observations \citep{2014AJ....148...75D,2018MNRAS.481.5247O}. These studies have focused on the radio properties, morphological characteristics, optical environments, and spatial correlations with galaxy clusters of BTRGs.
In recent years, BTRGs selected from the FIRST survey have been the subject of continuous research \citep{2019AA...626A...8M,2021ApJS..254...30P,2022ApJS..259...31S}, with \citet{2022ApJS..259...31S} notably reporting 717 BTRGs. Furthermore, the identification of BTRGs has expanded to encompass lower frequency bands, with significant contributions from the Tata Institute of Fundamental Research (TIFR) Giant Metrewave Radio Telescope (GMRT) Sky Survey Alternative Data Release 1 (TGSS ADR1) \citep{2022MNRAS.516..372B} and the first data release of the Low Frequency Array (LOFAR) Two-metre Sky Survey (LoTSS DR1) \citep{2021arXiv210315153P}.

The BTRG samples from the previous work mentioned above were all acquired via visual inspection. Using this labor-intensive approach alone results in a limited number of BTRG samples, and the subsequent statistical analysis of their physical properties cannot be automated, thereby consuming substantial manpower and time resources. Furthermore, to achieve a more profound understanding of the physical properties of BTRGs and their evolutionary processes, it is imperative to amass a significantly larger collection of BTRG samples. Recently, deep learning techniques have been increasingly applied to the automated morphological classification and detection of radio galaxies, as exemplified by studies such as those by \citet{2018MNRAS.476..246L,2018MNRAS.480.2085A,2019MNRAS.488.3358T,2019MNRAS.482.1211W,2019Galax...8....3L,2021SciBu..66.2145L,2021MNRAS.501.4579B,2021MNRAS.503.1828B,Zhang_2022,2023A&C....4200682R,2023A&C....4400728L,2024PASA...41....1G,2024PASA...41...27G}.  
Notably, the deep learning network described in \citet{2023A&C....4400728L} has been successfully applied to the FIRST survey, resulting in the compilation of a catalog comprising 45,241 straight FR-II radio galaxies \citep{2024RAA....24c5021L}. However, it is important to note that, until now and in the foreseeable future, achieving 100\% accuracy in object identification remains a challenge for artificial intelligence methods, including deep learning. Therefore, deep learning methods can quickly identify interesting candidates from massive datasets, saving us time. Visual inspection may then be used to evaluate those candidates in complex structure categories and eliminate those suspected of being misclassified, thus enhancing the reliability of the final results.

In this paper, we present the results of a thorough search employing a combination of deep learning and visual inspection methods for the identification of BTRGs from the latest release of the FIRST catalog. This paper is structured in the following sequence. In Section \ref{sect:Meth}, we provide a detailed description of the deep learning-based methodology and visual inspection process used to search and select BTRGs from the FIRST survey. In Section \ref{sec:hosts}, we present the methods of identifying host galaxies for our BTRGs and determining the redshifts of the corresponding host galaxies. The outcomes of our BTRG catalog, along with the calculations and result discussions pertaining to the physical properties, are detailed in Section \ref{sect:results}. The conclusion is presented in Section \ref{sec:cons}.

We adopt the $\Lambda$CDM cosmological model throughout this paper, with parameters specified as follows: $H_0 = 67.4 \,{\rm km}\,{\rm s}^{-1}\,{\rm Mpc}^{-1}$, $\Omega_m = 0.315$, and $\Omega_{vac}=0.685$ \citep{2020A&A...641A...6P}.

\section{Identifying BTRGs from the FIRST Survey Images} 
\label{sect:Meth}

\subsection{The FIRST Survey Data}%
The FIRST survey is a 1.4 GHz continuum survey that utilized the NRAO Karl G. Jansky VLA B configuration to carry out survey observations over 18 years (1993-2011), accumulating a total observation time of more than 4000 hours (5.5 months) \citep{1995ApJ...450..559B}. With an angular resolution of 5$''$ and an average rms noise level of 0.15 mJy, the FIRST survey covers 25\% of the entire sky, encompassing a total sky area of 10,575 ${\rm deg}^2$. This area is divided into 8444 ${\rm deg}^2$ in the north Galactic cap range of R.A. $7.0^{\rm h}$ to $17.5^{\rm h}$ and Decl. $-8^{\circ}.0$ to $+57^{\circ}.6$, and 2131 ${\rm deg}^2$ in the south Galactic cap range of R.A. $20.4^{\rm h}$ to $4.0^{\rm h}$, Decl. $-11^{\circ}.5$ to $+15^{\circ}.4$. In our BTRG identification, we used the December 17, 2014 version of the catalog (FIRST-14dec17) and accompanying images. The FIRST-14dec17 catalog contains approximately 946,432 components, generated by an Astronomical Image Processing System (AIPS) source extraction program HAPPY \citep{1997ApJ...475..479W} in the final FIRST images \citep{2015ApJ...801...26H}.

\subsection{RGCMT: A Deep Learning Based Radio Source Finder} \label{sec:finder}
\begin{figure*}[ht!]
\plotone{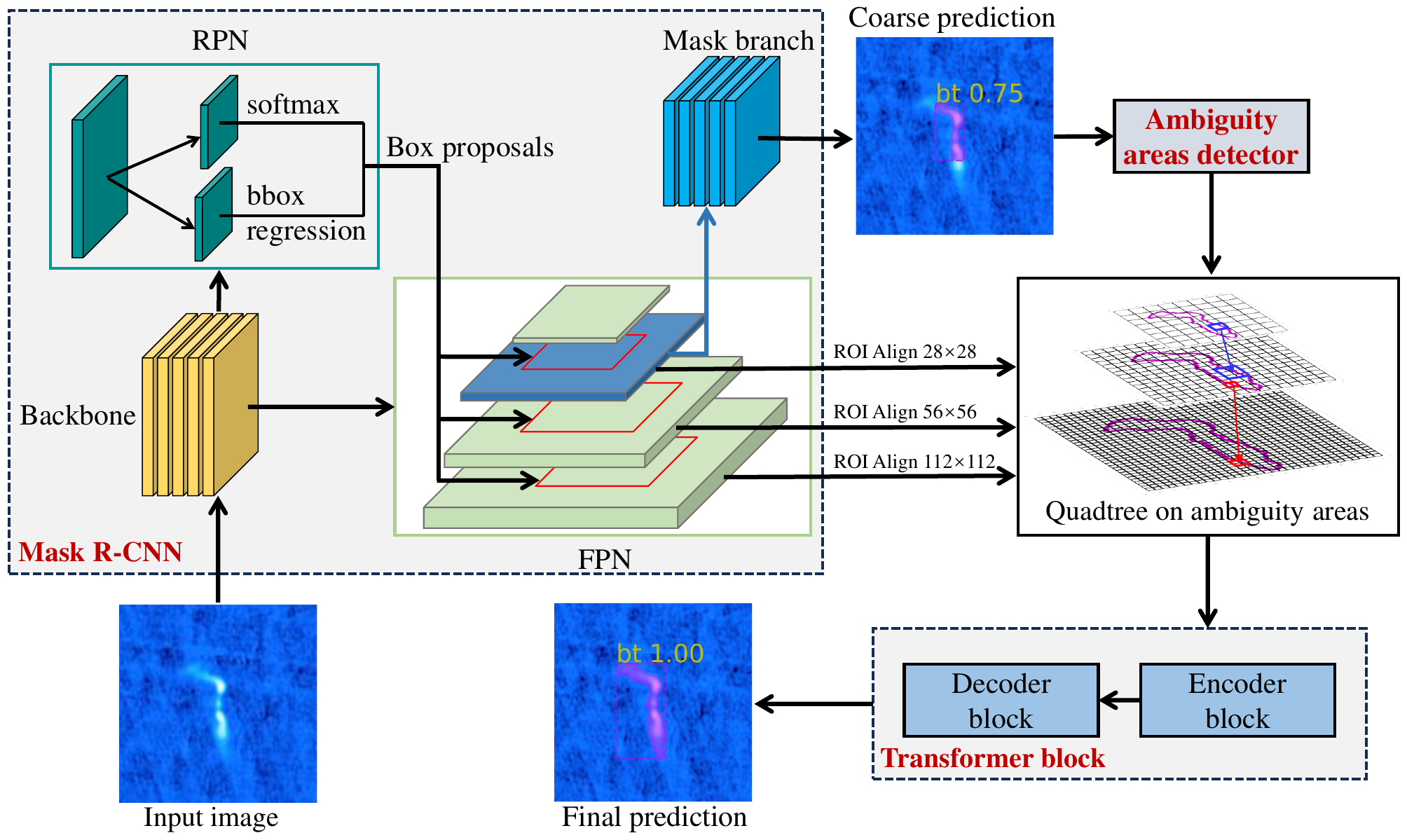}
\caption{A simple schematic of the RGCMT network for BTRG detection, modified from \citet{ke2022masktransfiner}. For a given image, Mask R-CNN first generates an initial coarse mask prediction. Then, the ambiguity areas detector identifies ambiguity areas using three levels of Region of Interest (RoI) Align features. Finally, the transformer block is used to correct the ambiguity areas and produce the final refined mask prediction. 
\label{fig:ht_finder}}
\end{figure*}

We use a deep learning based radio source finder called Radio Galaxies Classification with Mask Transfiner \citep[RGCMT;][]{2023A&C....4400728L}, which builds on the pioneering works of Mask Transfiner \citep{ke2022masktransfiner} and detectron2 \citep{wu2019detectron2}, to search for BTRGs from FIRST images. The overall architecture of RGCMT, as illustrated in Figure \ref{fig:ht_finder}, includes three primary parts: the Mask Region with Convolutional Neural Networks (CNN) \citep[Mask R-CNN;][]{he2017mask}, the ambiguity areas detector \citep{ke2022masktransfiner}, and the transformer block \citep{carion2020end}. For each detected object in an input image, RGCMT provides a predicted mask, bounding box, and score.

The Mask R-CNN serves as the base source detector. Initially, it utilizes the Residual Network \citep[ResNet;][]{he2016deep} backbone in conjunction with the Feature Pyramid Network \citep[FPN;][]{lin2017feature} to extract multi-scale feature maps from the input image. The Region Proposal Network \citep[RPN;][]{ren2015faster} then predicts bounding boxes as source proposals (box proposals). Subsequently, the multi-scale feature maps and box proposals are used to generate the Region of Interest (RoI) features pyramid. Finally, a coarse initial mask prediction is generated via the mask branch using a single-level aligned RoI (RoI Align) feature.     

The quality of mask prediction results directly impacts the positioning accuracy and morphology identification of radio sources \citep{2016MNRAS.460.1486R, 2019MNRAS.487.3971H}. Additionally, it also affects the statistical accuracy of subsequent physical properties of radio sources. In the initial mask predicted by Mask R-CNN, some pixel points in the object's boundary area are misclassified. This is because a single high-level feature obtained via the down-sampling in FPN is used, which results in the loss of the object's detailed information. We define these misclassified pixel points as ambiguity areas. To refine the mask prediction quality, a quadtree structure of ambiguity
areas detector is first used to identify the ambiguity areas across multi-scale RoI Align features. This ambiguity areas detector initially predicts ambiguity areas on a high-level (low resolution) RoI feature, and then the detected low-resolution masks are up-sampled and fused with a lower level (higher resolution) feature in adjacent scales to make finer predictions of ambiguity areas. One of the high level feature points has four quadrant points in its adjacent low-level feature map, forming a quadtree structure. Here, we use three levels of features: $28\times28$, $56\times56$, and $112\times112$ pixels. Finally, we refine the mask predictions in ambiguity areas using a transformer-based block comprising an encoder block and a decoder block \citep{carion2020end}. The use of multi-level features in fusion during mask prediction enables the retention of higher resolution and more detailed local information about the object. As a result, the RGCMT network achieves a more refined mask prediction result for the object.   

\begin{figure*}[ht!]
\plotone{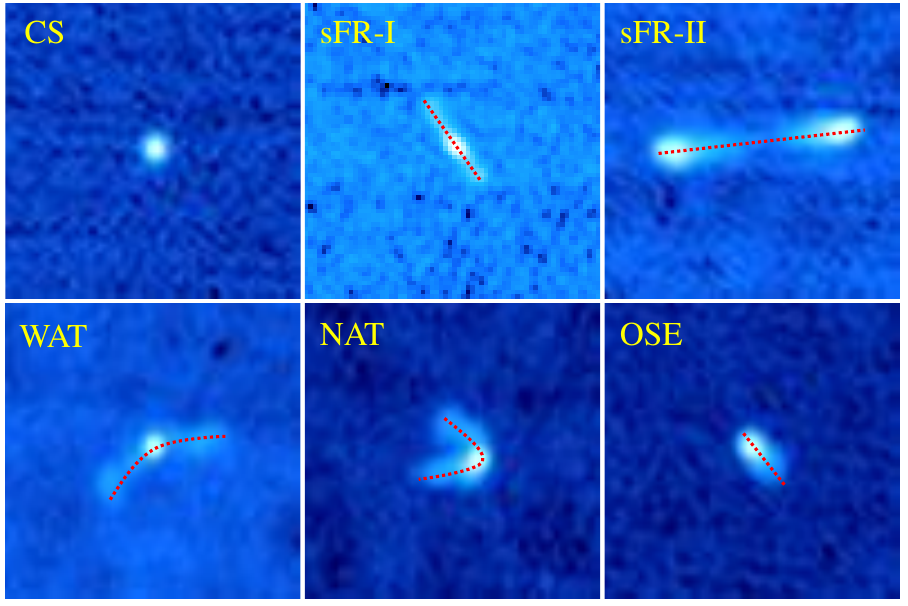}
\caption{Examples of five classes (CS, sFR-I, sFR-II, BT, and OSE) in this work. The BT class is a combination of WAT and NAT. The red dotted line represents the straight or bent jets of a radio source.
\label{fig:CL_SCHE}}
\end{figure*}

To distinguish BTRGs from other morphologies of radio galaxies in FIRST images, RGCMT classifies radio galaxies into two main categories and five classes. One category is bent jet sources, i.e., BT, while the other encompasses non-bent jets sources: non-extended diffuse emission sources or point-like sources, i.e., Compact Sources (CS), straight FR-I (sFR-I), straight FR-II (sFR-II), and One-sided Straight Extended diffuse sources (OSE). 
Example images of these five classes are shown in Figure \ref{fig:CL_SCHE}.  

We trained the RGCMT model using 3172 FIRST images, which contain 3601 sources. These sources consist of 1673 CSs, 326 sFR-Is, 738 sFR-IIs, 406 BTs, and 458 OSEs. Each image initially extracts its FITS file from the FIRST Cutout Server\footnote{\url{https://third.ucllnl.org/cgi-bin/firstcutout}} with an image size of 3$'$.96 ($132\times132$ pixels), and subsequently utilizes SAOImage DS9 \citep{2003ASPC..295..489J} to convert it to a three-channel (RGB) image in PNG format employing the cool color-map and the logarithmic min-max scale. The trained RGCMT model was evaluated on a labeled dataset consisting of 1946 images, including 894 CSs, 217 sFR-Is, 479 sFR-IIs, 270 BTs, and 304 OSEs. The data selection and labeling for the training and evaluation sets were mainly done by visual inspection according to the classification scheme in Figure \ref{fig:CL_SCHE} using the annotation tool Labelme \citep{russell2008labelme}. During this process, the class label of each source was determined by the consensus of three individuals. It was observed that the mean average precision (mAP) for the five classes, with an Intersection over Union (IoU) threshold of 0.5, is 98.4\% on the evaluation dataset. Here, the IoU represents the ratio of the area of intersection between the predicted mask and the ground truth mask of the detected source to the area of their union. The mAP and IoU, along with their related metrics, were calculated according to the Common Objects in Context evaluation metrics\footnote{\url{https://cocodataset.org/\#detection-eval}}. The average precision (AP) at the same IoU value for the CS, sFR-I, sFR-II, BT, and OSE classes is 97.3\%, 97.4\%, 99.9\%, 98.4\%, and 98.9\% respectively. This indicates that the RGCMT model exhibits high accuracy in detecting BTRG. For more details, please refer to \citet{2023A&C....4400728L}.

\subsection{BTRG Identification using RGCMT}
We utilized the trained RGCMT model to identify BTRGs in all unlabeled FIRST images. Each FITS image was obtained from the FIRST Cutout Server, with the image center positioned on the component as listed in the FIRST-14dec17 catalog. The same image size and pre-processing techniques described in Section \ref{sec:finder} were applied. After traversing and processing all components, a total of 946,432 images in FITS and PNG formats were obtained. The RGCMT model was used to sequentially process all images for the BTRG search. For each BTRG detection, the centroid and peak flux density positions (R.A. and Decl.) in J2000, total flux density, and local rms noise were calculated based on its predicted mask. After removing duplicate detections and excluding those with a confidence threshold (predicted score) less than 0.5 and a total flux density lower than 1.64 mJy, a total of 11,473 BTRG candidates were identified.

\subsection{Filtering Out Non-interesting Candidates using Visual Inspection}
The RGCMT model completed the detection of 946,432 images in 17.5 hours using 3 GPUs, achieving an average detection rate of 66.6 ms per image. This method significantly reduces the time required to find BTRGs compared to visually inspecting all images. Although the RGCMT model's BTRG detection boasts high accuracy, the results include candidates that are false positives and not real BTRGs (non-interesting candidates) due to the inherent limitations of artificial intelligence methods and the complex extended structure of BTRG. The information provided by the RGCMT model for each detected BTRG, such as bounding boxes, masks, and centroid positions, can accelerate or even automate post-processing tasks such as multi-band cross-identification and radio property statistics for BTRGs. Details can be found in Section \ref{sec:hosts} and Section \ref{sect:results}. 
To compile a highly reliable BTRG catalog, it is essential to meticulously filter out non-interesting candidates. Therefore, we visually inspected the FIRST contour images of each BTRG candidate detected by the RGCMT model and removed those candidates that belong to the following cases: 
(1) candidates that are not bent in the common direction, including X-/S-/Z-/W-shaped radio galaxies; (2) candidates that are lobes of large radio galaxies; (3) candidates that are actually artefacts arising from observation effects; (4) candidates that have $OA>170^\circ$ and do not have sufficient jet bending. 
Finally, 4876 out of 11,473 candidates have been retained to construct our final BTRG catalog.

\section{Finding Host Galaxies for BTRGs}\label{sec:hosts} 

\subsection{Host Galaxy Candidates}\label{sec:hosts_cand}
To comprehend the formation and evolution of BTRGs, it is imperative to systematically gather and analyze the physical properties of BTRG samples. However, radio sources, including BTRGs, typically lack detailed information beyond the spectral index in radio spectra. The most essential information (e.g. redshift, mass, and optical morphology) about radio sources needs to be obtained from observations at other wavelengths, such as optical and infrared (IR). From this information, we can also determine other physical properties of the radio sources. For example, if the redshift is known, the radio luminosity can be determined. Therefore, the primary task is to search for the counterparts of BTRGs, namely host galaxies, in optical or IR wavelengths.
Generally, the host galaxy of a BTRG is located close to the peak of brightness or the regional center of the radio galaxy \citep{2022MNRAS.516..372B,2022ApJS..259...31S}.
We identify potential host galaxy candidates for our BTRGs from the Dark Energy Spectroscopic Instrument (DESI) Legacy Surveys (LS) \citep[DESI LS;][]{2019AJ....157..168D} using the finding-all-match method.

The DESI LS provides superior multi-color imaging with improved depth and resolution compared to earlier surveys like the Sloan Digital Sky Survey \citep[SDSS;][]{2000AJ....120.1579Y,2022ApJS..259...35A} or the first part of the Panoramic Survey Telescope \& Rapid Response System \citep[Pan-STARRS1;][]{2020ApJS..251....7F}. The sky area containing all BTRG samples in this study is included in the DESI LS's tenth data release (DR10). This release extends photometric coverage to over 20,000 ${\rm deg}^2$ and features new $i$-band images from Dark Energy Camera (DECam) data by the National Optical-Infrared Astronomy Research Laboratory (NOIRLab). It also integrates $grz$ images from three initial surveys (the Beijing-Arizona Sky Survey \citep[BASS;][]{2017PASP..129f4101Z}, the DECam Legacy Survey \citep[DECaLS;][]{2019AJ....157..168D}, and the Mayall z-band Legacy Survey \citep[MzLS;][]{2019AJ....157..168D}) and four infrared bands from Wide-field Infrared Survey Explorer (WISE) and Near-Earth Object WISE (NEOWISE).
Hence, we utilize the DESI LS DR10, querying sources from it within 30$''$ of the BTRGs' peak flux density coordinates and centroid coordinates using the X-Match Service at Astro Data Lab \citep{2014SPIE.9149E..1TF,2020A&C....3300411N}.

The method described above identifies one or more potential host galaxy candidates for each BTRG. A total of 211,910 optical candidates were identified for our BTRGs. Generally, the candidate with the smallest separation from
the location of the peak brightness or the
centroid positions of the BTRG is considered the actual host for a BTRG with a simple structure and well-aligned in radio and optical wavelengths. However, BTRGs often exhibit complex extended structures, and the candidate with the smallest separation may not always be the true host. The host candidates require further verification. Many alternative automated methods, such as Likelihood Ratio \citep{1992MNRAS.259..413S,2012MNRAS.423..132M}, have been introduced. Nevertheless, these methods are limited to point-like sources or simple extended sources, such as double radio sources \citep{2018MNRAS.473.4523W,2023ApJS..267...37G}. Additionally, a multi-wavelength cross-identification method for extended radio sources utilizing machine learning has also been proposed \citep{2018MNRAS.478.5547A,2022MNRAS.516.4716A}, but the accuracy of this method is not as high as desired and thus needs to be further improved. Therefore, this work employs visual inspection to further validate the identified candidates.

\subsection{Identifications from Visual Inspection}\label{sec:hosts_vi}
\begin{figure*}[!ht]
\centering
\includegraphics[scale=0.5]{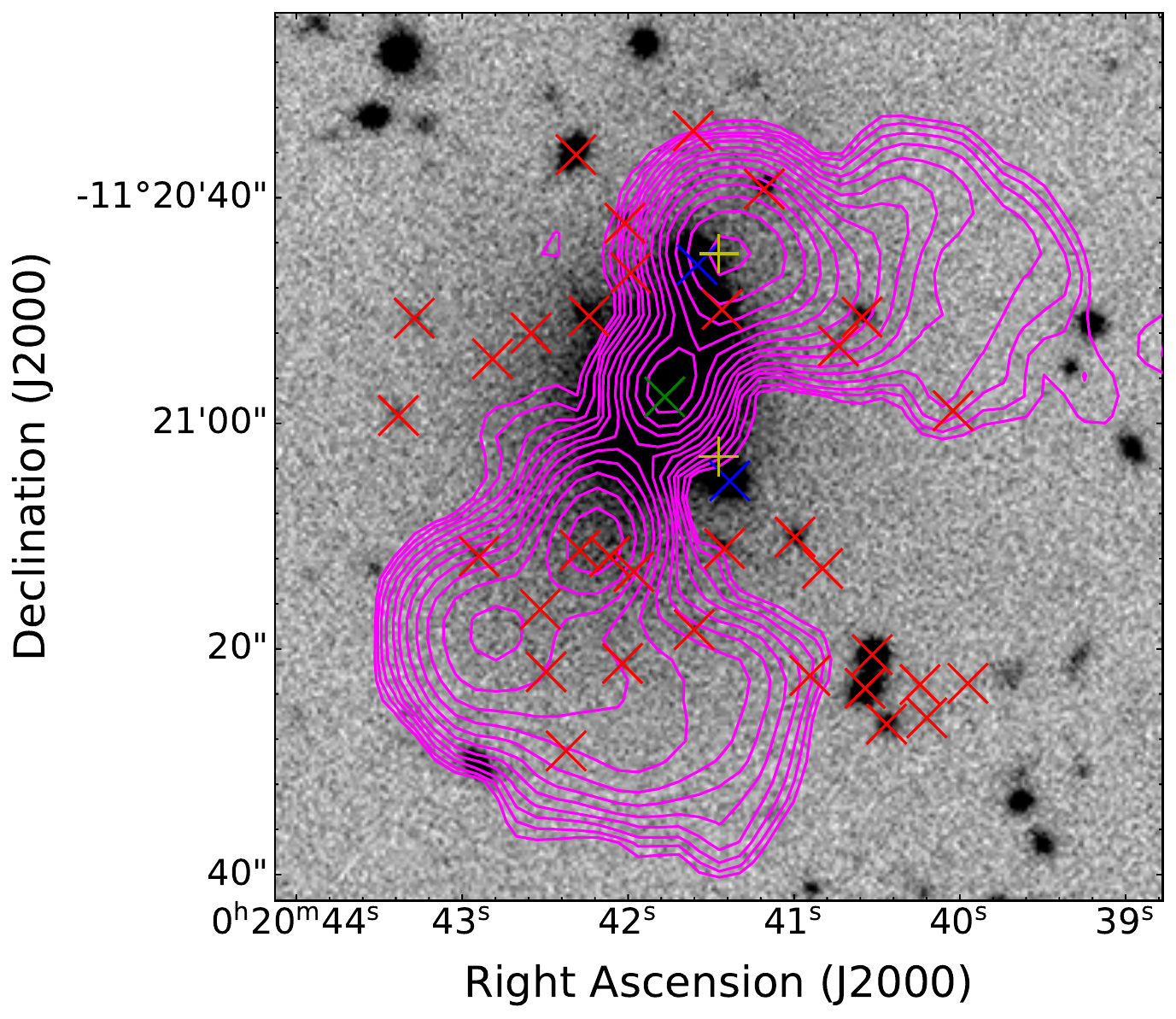}
\caption{An example demonstrating the identification of a host galaxy through visual inspection of a FIRST image (magenta contours) overlaid on a DESI LS $r$-band image (background image). Contour levels begin at 3 times the local rms noise (1$\sigma$ =0.14 mJy/beam) and increase by a factor of $\sqrt{2}$. 
The peak flux density and centroid coordinates are marked with a yellow `+'. Potential host candidates are indicated with an `$\times$', where a green `$\times$' represents the true host galaxy identified by visual inspection, and two blue `$\times$' signs indicate the nearest host candidates to the peak flux density and centroid coordinates, respectively. 
\label{fig:VI_exam}}
\end{figure*}

We overlay FIRST contours of BTRGs onto their corresponding optical $r$-band images from DESI LS DR10, marking the potential host candidates' coordinates on the images with an `$\times$'. The pixel size of each image is slightly larger in width and height than that of the predicted bounding box of the BTRG sample. We then use visual inspection to check these images one by one to determine the final correct host, if any. An example of a FIRST and DESI LS map of a BTRG sample is shown in Figure \ref{fig:VI_exam}. 

Figure \ref{fig:VI_exam} shows that this BTRG has 36 host candidates identified by the method described in Section \ref{sec:hosts_cand} and marked with an `$\times$’. Through a discerning visual examination, we initially dismissed the 33 host candidates indicated by red `$\times$’ markers, as they either did not reside within the radio-emitting region pertaining to this BTRG or exhibited a lower SNR in the $r$-band. Next, despite the host candidate adorned with two blue `$\times$' being spatially proximate to the peak flux and centroid coordinates, respectively, its alignment with the radio core proved to be incongruous. In contrast, the host candidate distinguished by a green `$\times$’ coincides with the radio core. Consequently, we confidently designate this candidate as the true host. Additionally, after obtaining the hosts through visual observation, we further examined their types carefully and excluded those not of the `Galaxy' or quasi-stellar object (QSO) types. Utilizing this meticulous approach, we have successfully identified host galaxies for 4193 out of 4876 BTRGs. For the remaining BTRGs, no hosts were found because there were no reasonable candidates from DESI LS DR10.

\subsection{Redshifts} \label{sec:hosts_z}
Redshift assumes a critical role in the statistical analysis of radio galaxies, serving as a gateway to a wealth of physical data. Upon the precise determination of a radio galaxy's redshift, a trove of physical information is unveiled, such as luminosity distance, radio luminosity, projected largest linear size, etc. Using the host positions identified from DESI LS DR10, we first find spectroscopic redshifts from the NASA/IPAC Extragalactic Database (NED)\footnote{\url{https://ned.ipac.caltech.edu/}} for our BTRGs with a search radius of 1$''$. We select the spectroscopic redshifts that are nearest to the host positions of the BTRGs. 

For hosts with no spectroscopic redshift found in the NED, we then queried spectroscopic redshifts for them from a number of catalogs rich in spectroscopic data within a 1$''$ radius. 
The spectroscopic catalogs used are the 16th Data Release of the Sloan Digital Sky Surveys \citep[SDSS DR16;][]{2020ApJS..249....3A}, the data release 4 (DR4) of the Galaxy And Mass Assembly (GAMA) survey \citep[GAMA DR4;][]{2022MNRAS.513..439D}, the 2dF Galaxy Redshift Survey \citep[2dFGRS;][]{2001MNRAS.328.1039C}, the 6dF Galaxy Survey \citep[6dFGS;][]{2009MNRAS.399..683J}, the WiggleZ Dark Energy Survey \citep{2018MNRAS.474.4151D}, and the Two Micron All Sky Survey \citep[2MASS;][]{2006AJ....131.1163S} Redshift Survey \citep[2MRS;][]{2012ApJS..199...26H}. 

Not all BTRGs have spectroscopic redshifts. As a result, photometric redshift measurements often act as a supplementary means to enrich the redshift dataset. For the remaining hosts with an absence of redshift identifications, we searched for photometric redshifts within the DESI LS DR10 using the query interface provided by the Astro Data Lab \citep{2014SPIE.9149E..1TF,2020A&C....3300411N}.  
Finally, our meticulous search yielded redshifts for 4171 BTRGs, comprising 2357 spectroscopic redshifts (2151 from NED, 152 from SDSS DR16, 3 from GAMA DR4, 6 from 2dFGRS, 9 from 6dFGS, 15 from WiggleZ, and 21 from 2MRS) and 1814 photometric redshifts (all from DESI LS DR10).

\section{Results and Discussion}
\label{sect:results}
\subsection{The New BTRG Catalog}
We report a new BTRG catalog (BTRGcat) compiled from the FIRST survey, which contains a total of 4876 BTRGs, as listed in Table \ref{table:TRGcat}. Column 1 gives the names of the BTRGs derived from their radio centroid positions in the FIRST image provided by the RGCMT model. Column 2 lists the host names of BTRGs. For those BTRGs with available spectroscopic redshifts, we used the exact optical or IR names from NED, SDSS DR16, GAMA DR4, 2dFGRS, 6dFGS, WiggleZ Dark Energy Survey or 2MRS; for the remaining BTRGs, we used the DESI names. Columns 3 and 4 provide the coordinates of the host galaxies. For those without hosts, the coordinates of the radio name in column 1 are given. Column 5 presents the redshift values, accompanied by a flag type to indicate whether they are photometric or spectroscopic redshifts. The redshifts of all BTRGs with measured values span a range from 0.0023 to 3.43.

Column 6 lists the 1.4 GHz flux densities of BTRGs, which were measured from NVSS maps instead of FIRST maps, with values ranging from 3.81 mJy to 140.82 Jy. This choice is due to the superior accuracy of NVSS data in measuring fluxes for extended sources and its ability to detect low-surface-brightness objects that may be overlooked by FIRST data \citep{2022ApJS..259...31S,2024RAA....24c5021L}. Column 7 lists the 3 GHz flux densities of BTRGs calculated from available VLASS data. Column 8 presents the two-point spectral indexes ($\alpha_{1.4{\rm GHz}}^{3{\rm GHz}}$) between 1.4 GHz and 3 GHz for sources with available VLASS flux densities (see more details in Section \ref{sec:spec_index}). In column 9, we have also calculated the radio luminosity of the BTRGs at 1.4 GHz for those with known redshifts. For further details, please refer to Section \ref{sec:Lrad}. Columns 10 and 11 list the $OA$ and radius of curvature of the radio jets ($R_c$), respectively. For a comprehensive discussion on the calculation of $OA$ and $R_c$, refer to Section \ref{sec:oa_rpa}.

Columns 12 and 13 list the largest angular size (LAS) and largest linear size (LLS) of our BTRG samples. The LAS was determined by measuring the largest distance between any two points of the predicted polygon of the BTRG detected in the FIRST map. Columns 14 and 15 present the absolute $r$-band magnitude ($M_r$) and the logarithm of the black hole mass ($M_{\rm BH}$), which were estimated from the SDSS DR16 database. The $M_r$ values were directly queried as `absMagR' from the Photoz table. For $M_{\rm BH}$ values, we first queried the stellar velocity dispersion ($\sigma$) from the spectral data in SDSS DR16, and then computed them using the well-established $M_{\rm BH}-\sigma$ relation \citep{2002ApJ...574..740T}. Columns 16 and 17 list the $g$-$r$ color values and the morphological types of the BTRGs. The BTRGs were determined to be either FR-I or FR-II by visual inspection according to the standard FR classification scheme \citep{1974MNRAS.167P..31F}. For those BTRGs not conforming to the FR class definition, they were not classified as FR-I or FR-II. Among all BTRGs, 1221 (25\%) are FR-I type, 1799 (37\%) are FR-II type. 
Subsequently, according to the value of the $OA$, BTRGs with an $OA$ larger than 90$^\circ$ are determined to be WAT, while those with an $OA$ smaller than 90$^{\circ}$ are determined to be NAT.

\begin{figure*}[!ht]
\centering
\includegraphics[scale=0.25]{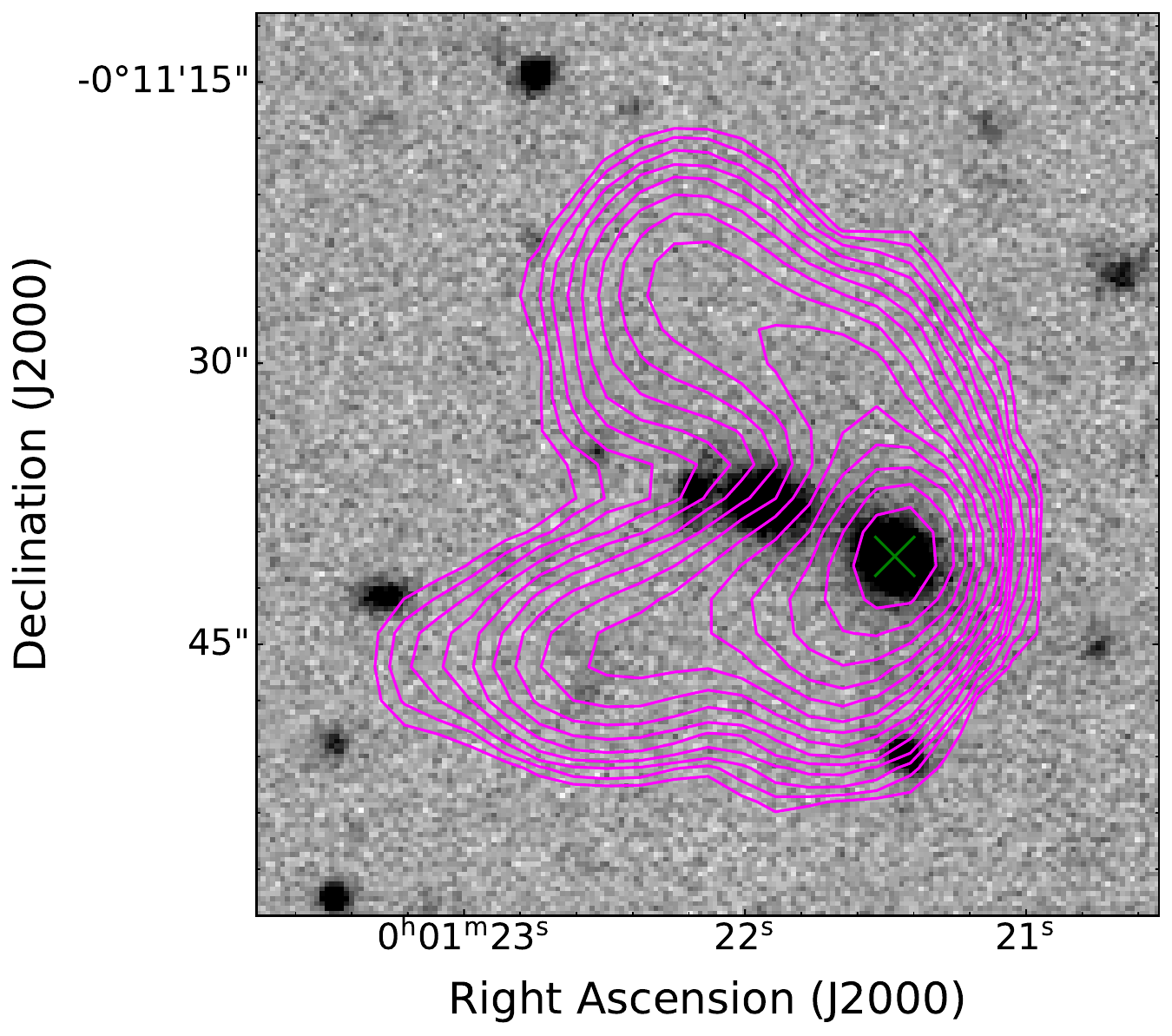}
\includegraphics[scale=0.25]{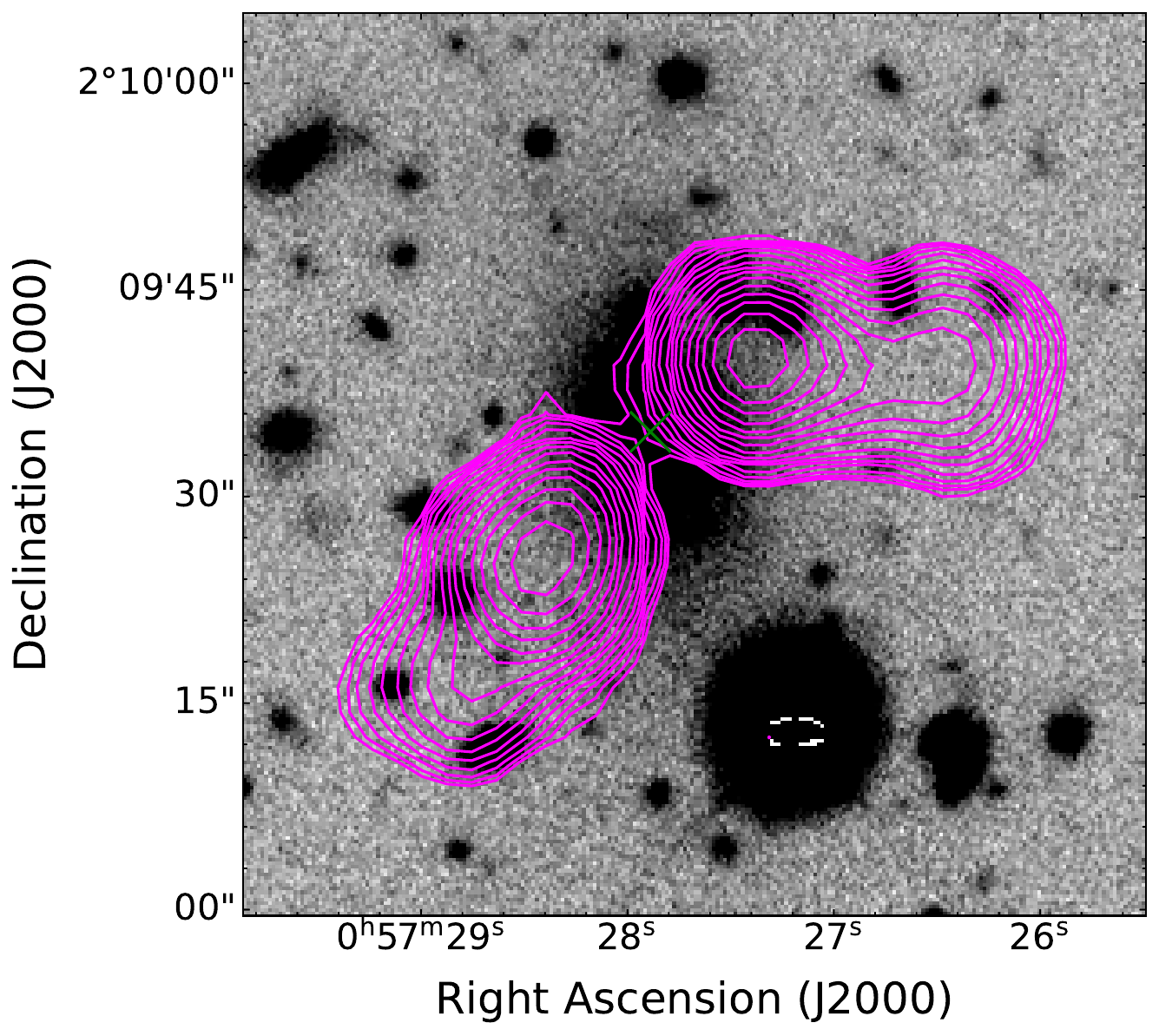}
\includegraphics[scale=0.25]{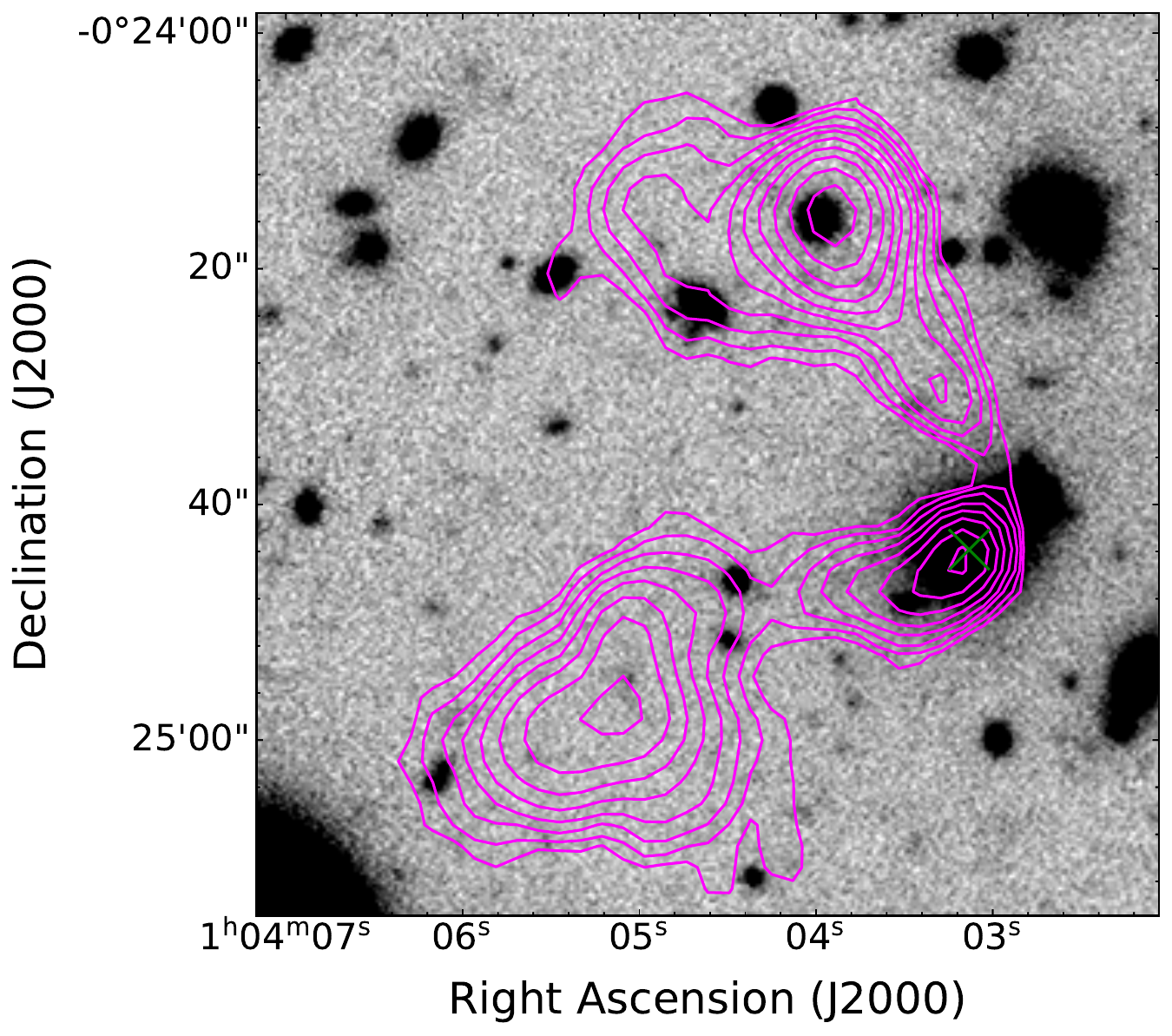}
\hspace{20mm}
\includegraphics[scale=0.25]{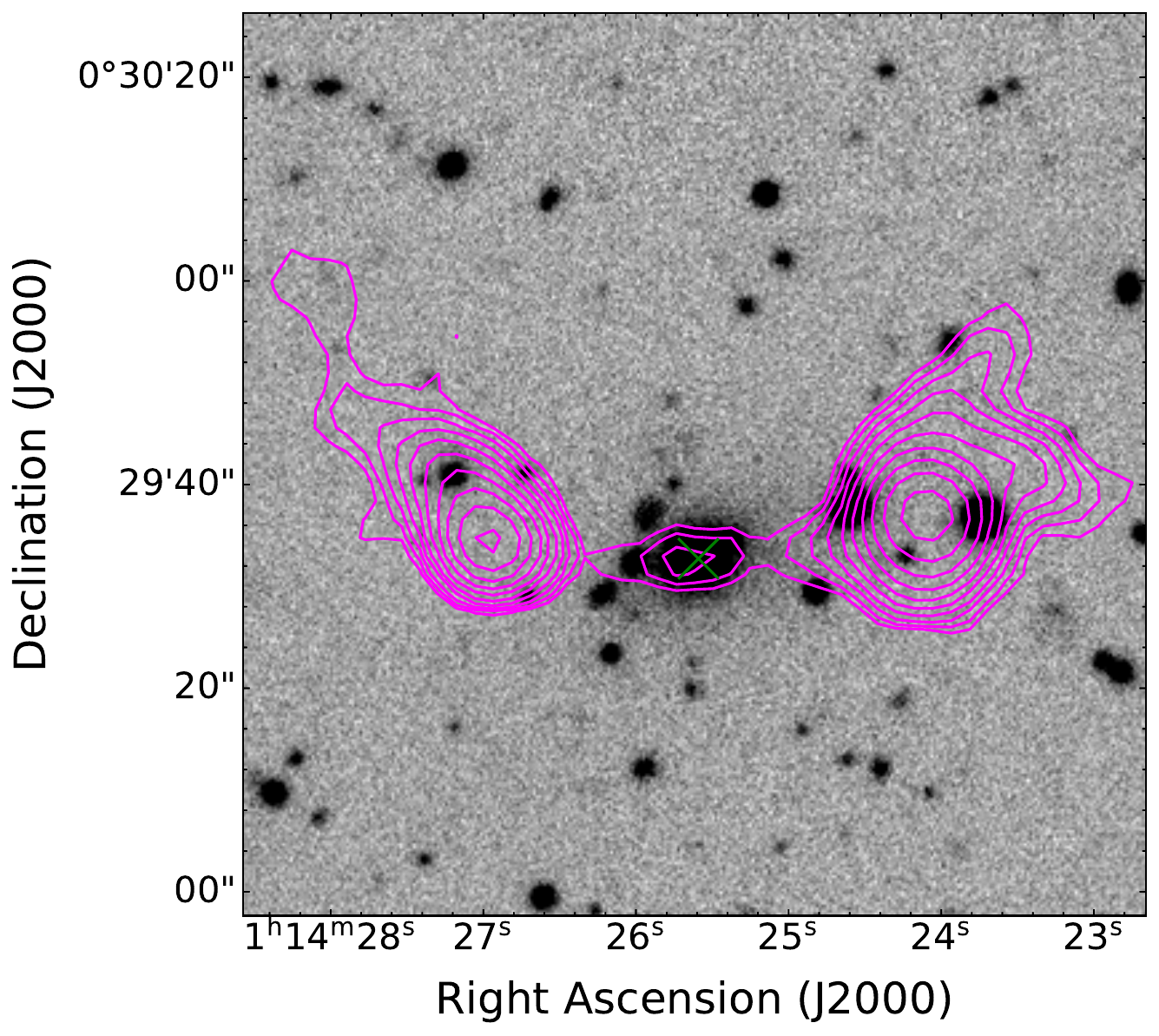}
\includegraphics[scale=0.25]{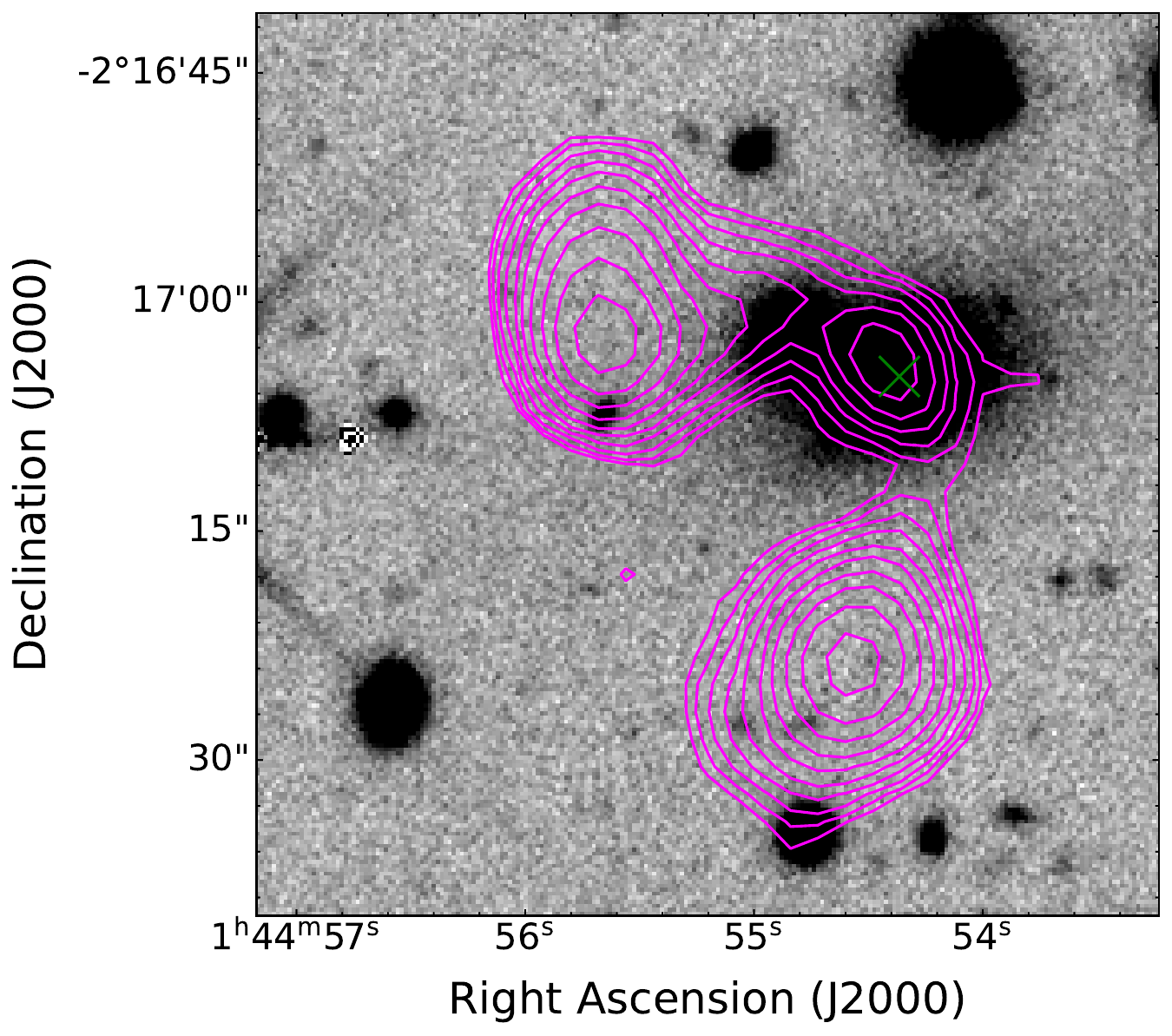}
\includegraphics[scale=0.25]{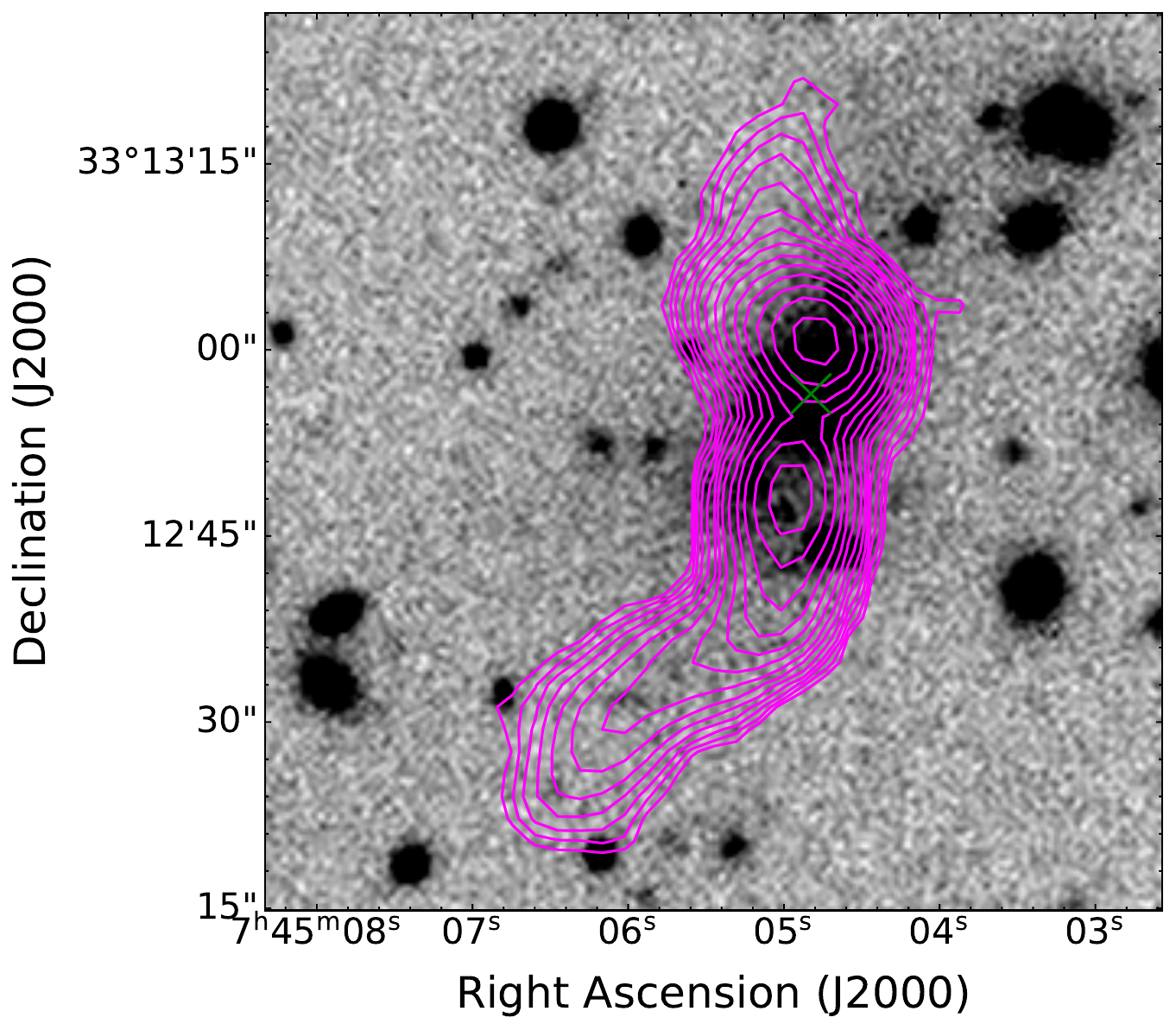}
\hspace{20mm}
\includegraphics[scale=0.25]{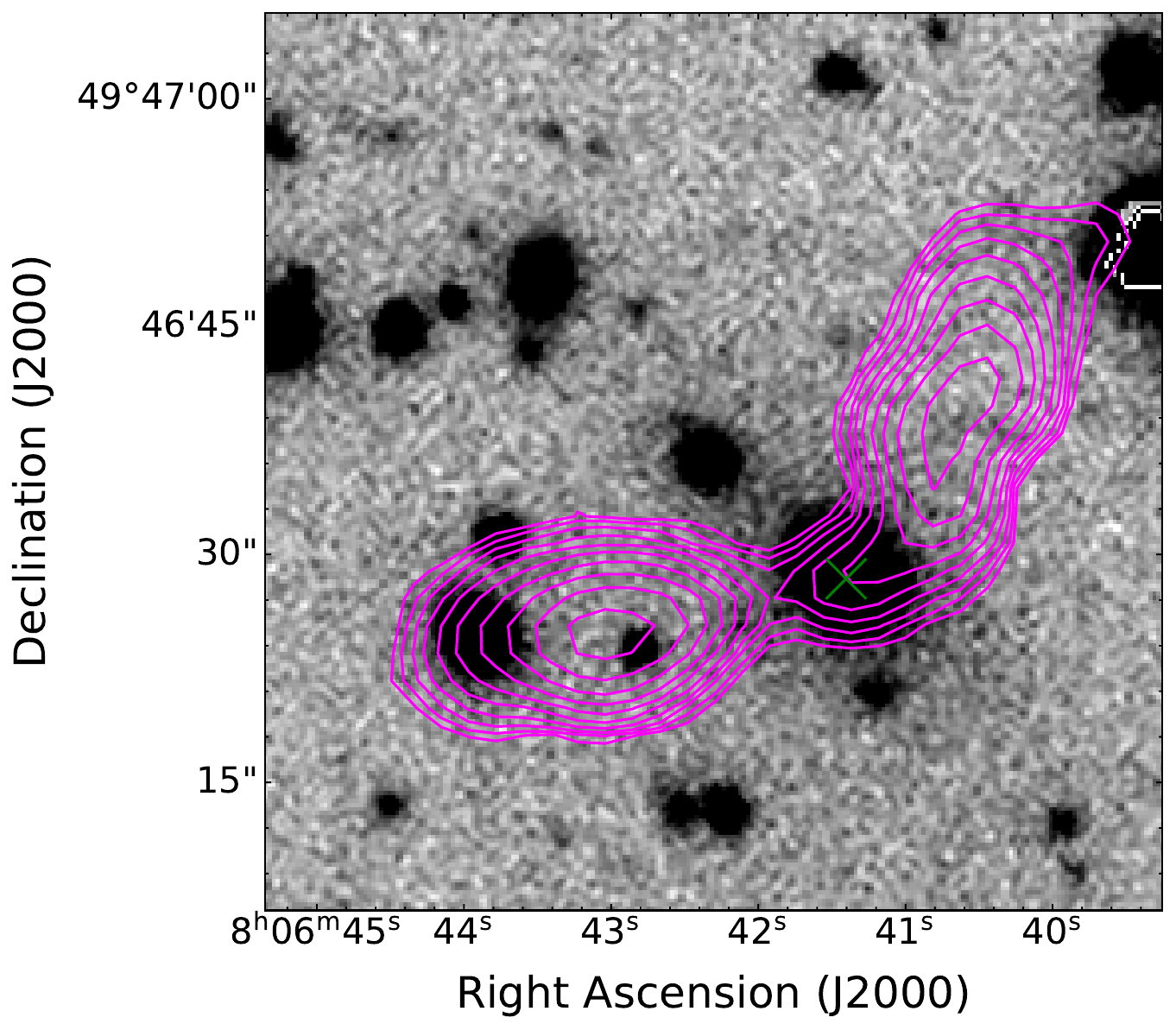}
\includegraphics[scale=0.25]{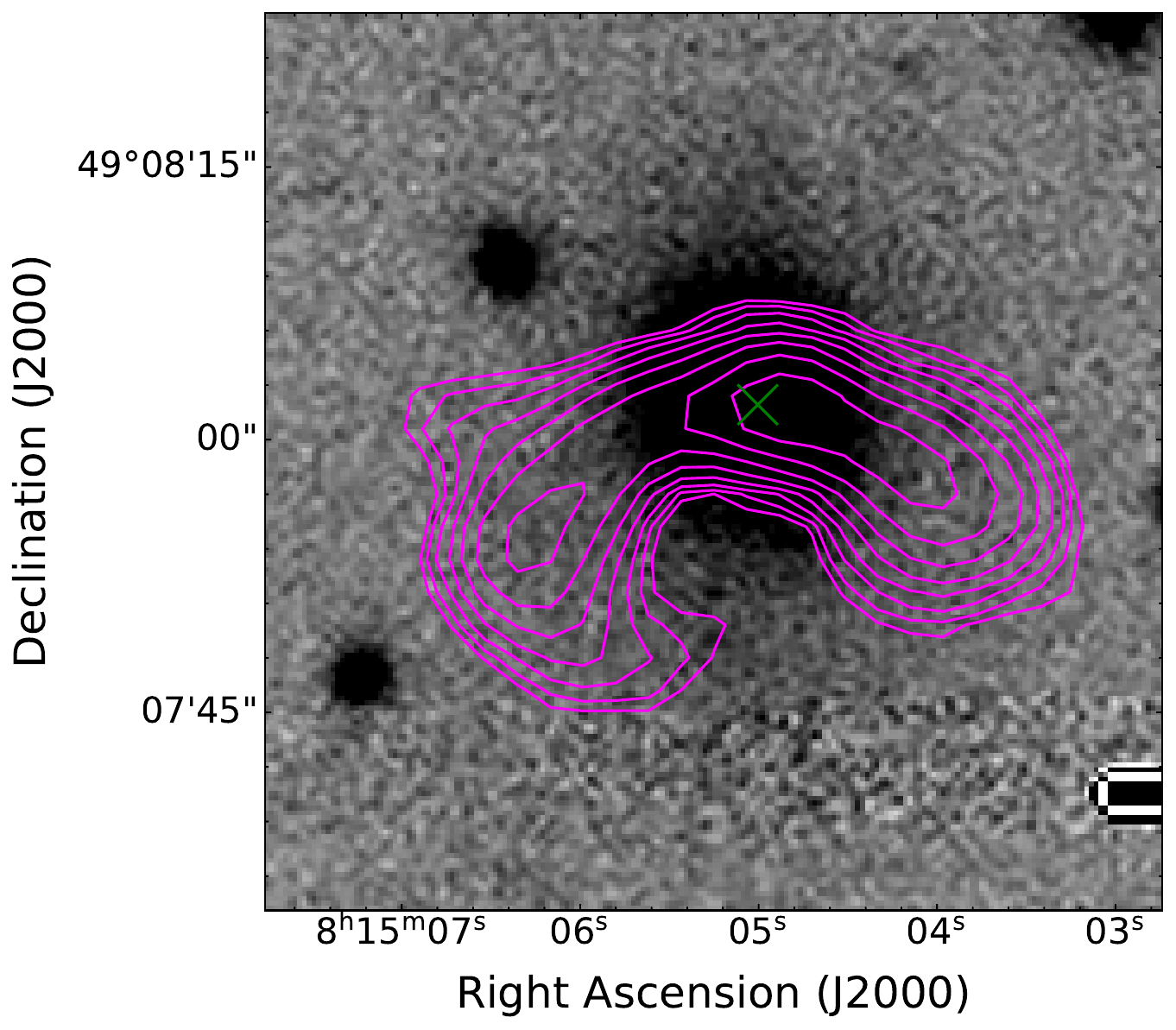}
\includegraphics[scale=0.25]{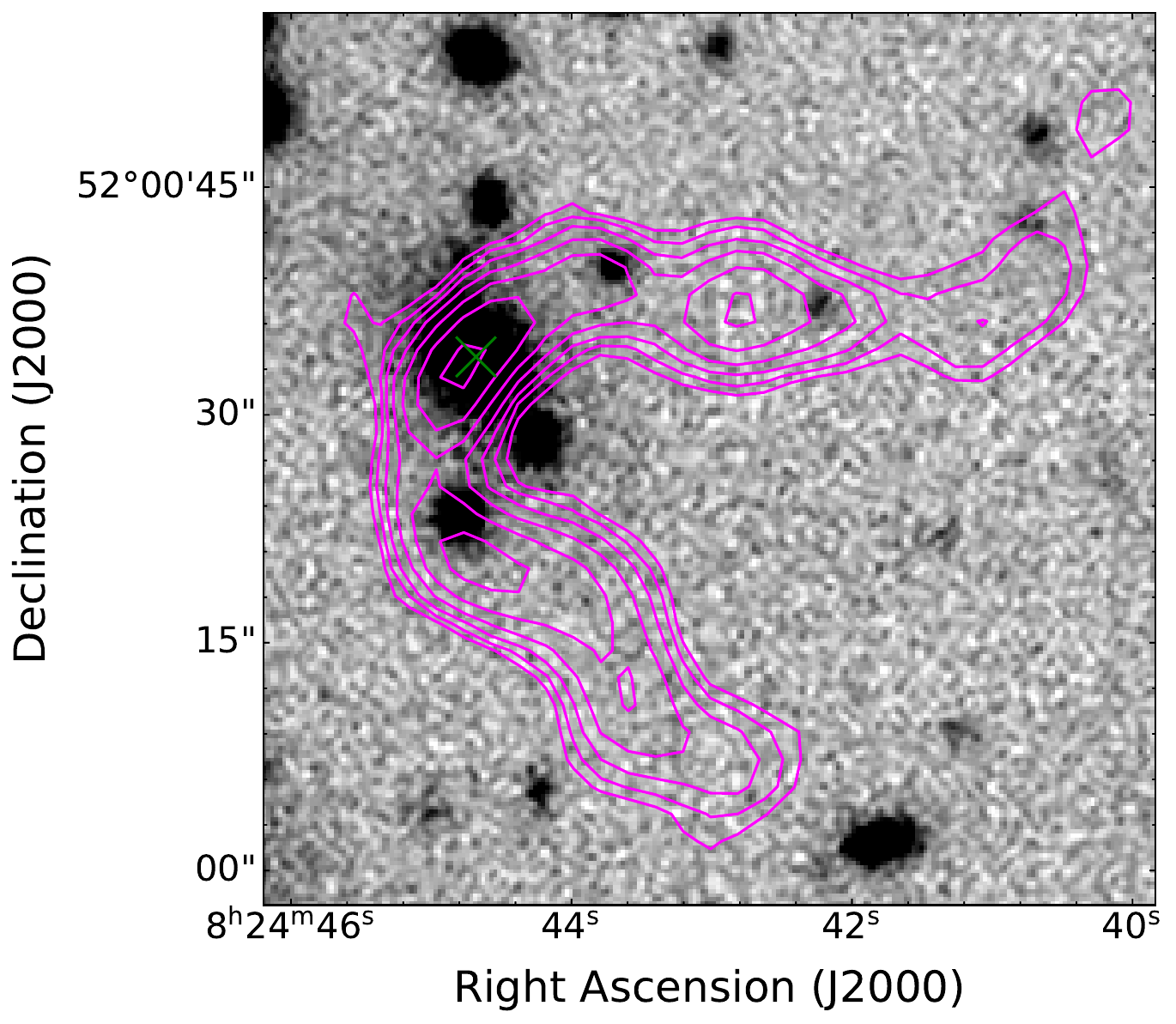}
\hspace{20mm}
\includegraphics[scale=0.25]{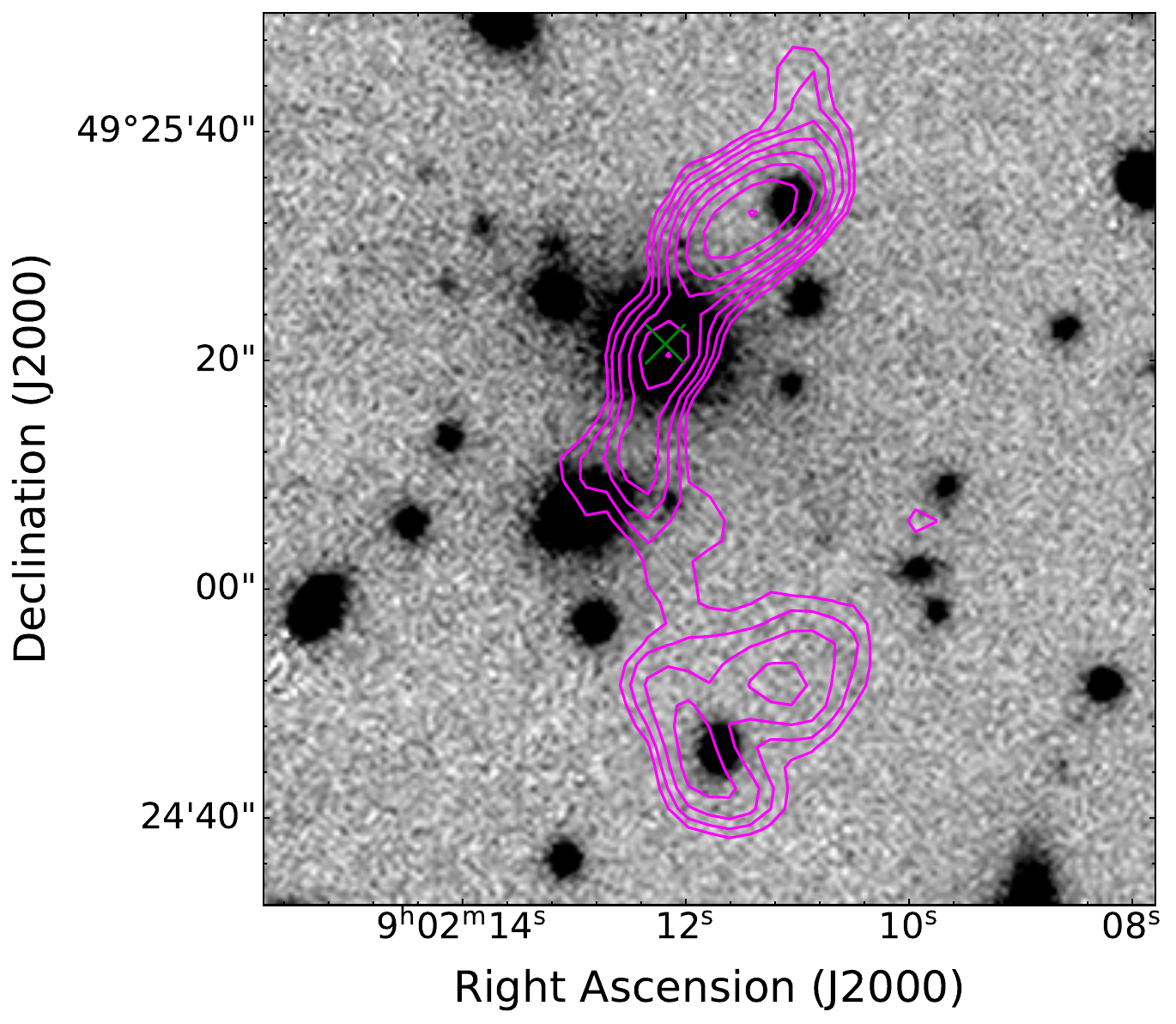} 
\includegraphics[scale=0.25]{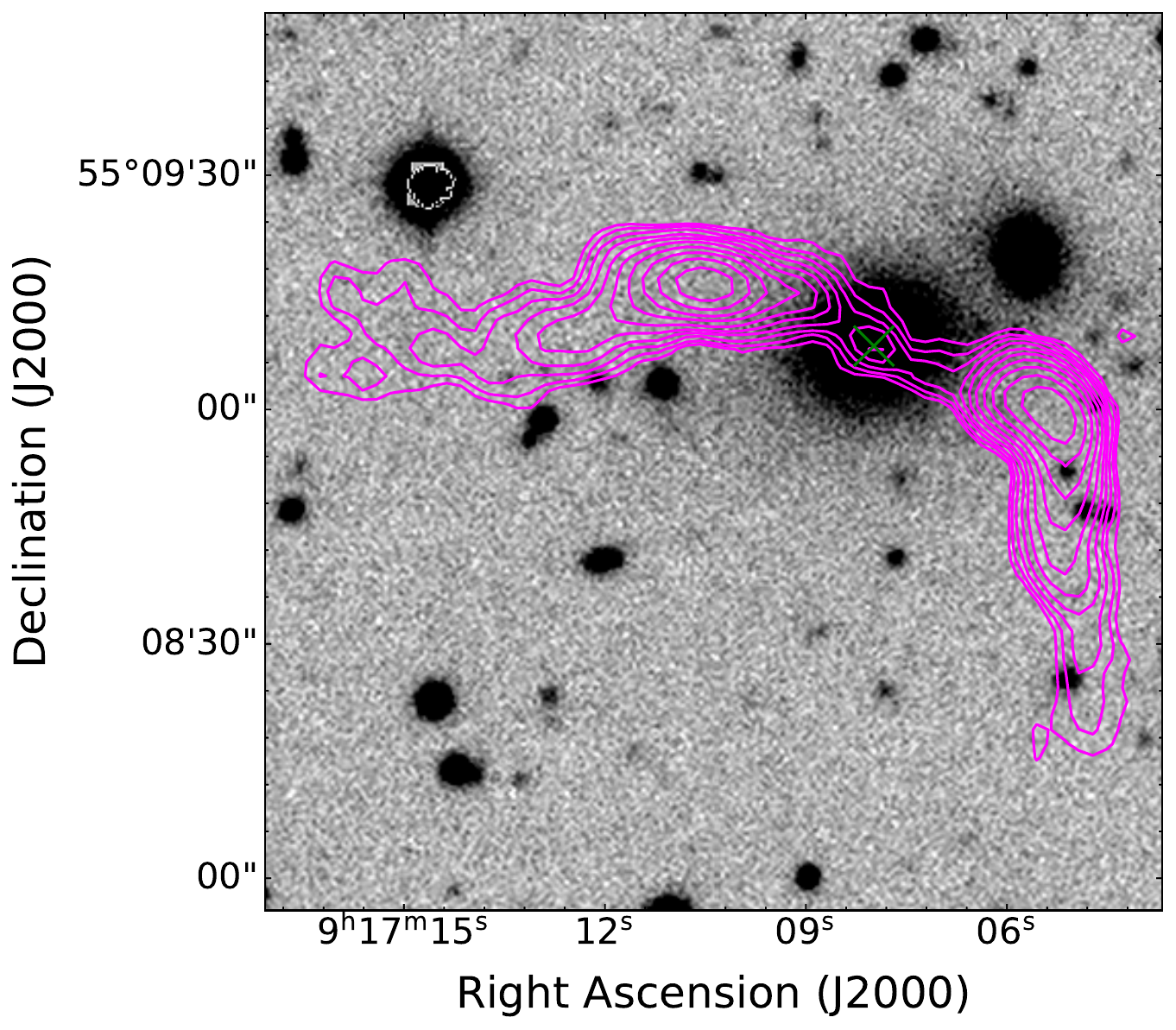} 
\includegraphics[scale=0.25]{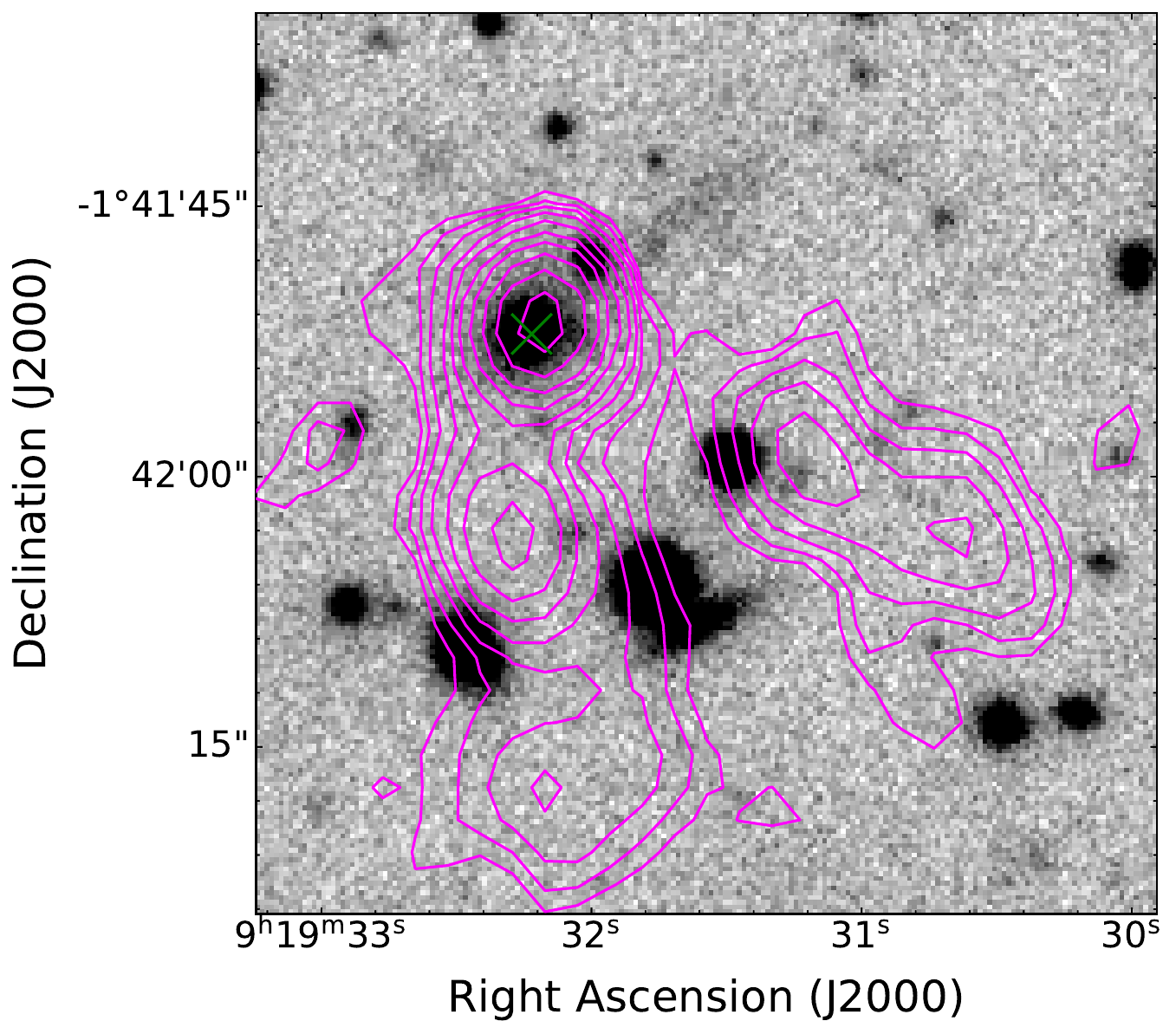}
\caption{FIRST image (contours) overlaid on the DESI LS r-band image (gray scale) for a sample of 12 BTRGs. Contour levels begin at 3 times the local rms noise and increase by a factor of $\sqrt{2}$. The green `$\times$’ represents the host galaxy. 
\label{fig:TRG_sample}}
\end{figure*}

Among the 4876 BTRGs in our catalog, 156 belong to the sample of \citet{2011AJ....141...88W} (marked as `a'), 29 are included in the sample of \citet{2011ApJS..194...31P} (marked as `b'), 193 are part of the sample from \citet{2017ApJ...844...78P} (marked as `c'), 36 are within the sample of \citet{2019AA...626A...8M} (marked as `d'), one is in the sample of \citet{2021ApJS..254...30P} (marked as `e'), 506 are included in the sample of \citet{2022ApJS..259...31S} (marked as `f'), 73 belong to the sample of \citet{2017ApJS..230...20A} (marked as `g'), 37 are in the sample of \citet{2017MNRAS.466.4346M} (marked as `h'), and 241 are included in the sample of \citet{2023DIB....4708974G} (marked as `j'), as detailed in Table \ref{table:TRGcat}. These nine samples have many BTRGs in common. After de-duplication, there are 1005 known BTRGs in all. Consequently, a total of 3871 BTRGs are newly discovered in our catalog. Figure \ref{fig:TRG_sample} shows radio-optical overlays of 12 examples of newly discovered BTRGs. We investigated possible reasons why our BTRG catalog does not fully contain all BTRGs in the latest BTRG catalog from \citet{2022ApJS..259...31S}. We found that the sources are present in \citet{2022ApJS..259...31S} but absent from our catalog because they are excluded by our selection criteria. First of all, sources with an $OA$ greater than 170$^\circ$ were screened out. For instance, source J0044$+$1026 in \citet{2022ApJS..259...31S} is an sFR-II.
Secondly, straight one-sided extended sources were eliminated, like sources J1321$-$0637 and J1521$+$5104 in \citet{2022ApJS..259...31S}.
Thirdly, sources with bends in different directions were also excluded; for example, J1138$+$2039 in \citet{2022ApJS..259...31S} is an S-shaped source. 


\movetabledown=172mm
\begin{rotatetable*}
\begin{deluxetable}{lcccccccccccccccc}
\tabletypesize{\scriptsize}
\tablecaption{A catalog of 4876 BTRGs identified from VLA FIRST survey\label{table:TRGcat}}
\tablehead{
\colhead{Radio name} &  \colhead{Host name}   & \colhead{R.A.} & \colhead{Decl.}  & \colhead{Redshift} & \colhead{$F_{1.4\,{\rm GHz}}$} & \colhead{$F_{3\,\rm GHz}$} & \colhead{$\alpha_{1.4\,{\rm GHz}}^{3\,{\rm GHz}}$} &  \colhead{${\rm log_{10}}$($L_{1.4\,{\rm GHz}}$)} & \colhead{$OA$}   & \colhead{$R_c$}  & \colhead{LAS} & \colhead{LLS} & \colhead{$M_r$} & \colhead{${\rm log_{10}}(M_{\rm BH})$} & \colhead{g-r} & \colhead{Type}     \\
 \colhead{} & \colhead{} &\colhead{(deg)} & \colhead{(deg)}  & \colhead{($z$)} & \colhead{(mJy)} & \colhead{(mJy)} & \colhead{} & \colhead{${\rm{(W}}\,{{\rm Hz}^{ - 1}})$} & \colhead{(deg)} &  \colhead{(arcsec)} & \colhead{(arcsec)} & \colhead{(kpc)} &\colhead{(mag)} &\colhead{($M_\odot$)} & \colhead{(mag)} & \colhead{}
}
\colnumbers
\startdata
J000111.57$-$002012.0 & DESI J000111.20$-$002011.6 &0.29668 &$-$0.33655 &0.52\tablenotemark{s} &66.89 &54.34 &0.27 &25.74 &133.0 &33.4 &41 & 263 &-- &-- &0.12 &FR-I,WAT \\
J000115.37$-$082639.6\tablenotemark{b,f} &SDSS J000115.11$-$082646.2 &0.31298 &$-$8.44618 &0.33\tablenotemark{s} &154.89 &75.44 &0.94 &25.76 &48.2 &7.9 &40 &196 &$-$23.17 &8.60 &1.68 &FR-II,NAT \\
J000121.53$+$010147.4 &WISEA J000121.52$+$010149.3 &0.3397 &1.03037 &0.55\tablenotemark{s} &17.27 &13.91 &0.28 &25.21 &161.9 &30.8 &34 &225 &$-$23.60 &8.12 & 1.85&--,WAT \\
J000122.13$-$001135.4 &DESI J000121.47$-$001140.3 &0.33959 &$-$0.19445 &0.46\tablenotemark{s} &118.45 &70.76 &0.68 &25.95 &72.8 &12.7 &36 &217 &$-$22.09 &8.35 &0.62 &FR-I,NAT \\
J000331.85$+$002812.0\tablenotemark{j} &WISEA J000330.70$+$002756.8 &0.87794 &0.46578 &0.19\tablenotemark{s} &43.38 &18.48 &1.12 &24.70 &112.8 &38.4 &76 &251 &$-$22.95 &9.06 & 1.28 &FR-I,WAT \\
J000353.42$+$121031.1 &WISEA J000353.27$+$121024.0 &0.97197 &12.17334 &0.76\tablenotemark{s} &241.04 &132.36 &0.79 &26.79 &98.1 &21.1 &34 &258 &$-$23.15 &8.45 &0.95 &FR-I,WAT \\
J000436.48$+$104724.0	&DESI J000436.75$+$104717.1	&1.15311	&10.78809	&0.58\tablenotemark{p}	&27.01	&9.53	&1.37	&25.67	&168.4	&269.2	&58	&393 &--&-- & 2.12 &FR-I,WAT\\
J000443.77$+$062411.4 & WISEA J000443.72$+$062404.7 &1.18221 &6.40131 &0.55\tablenotemark{s} &23.31 &11.96 &0.88 &25.46 &89.3 &8.7 &21 &139 &-23.75 &8.75 &1.93 &FR-I,NAT \\
J000511.32$-$075556.9	&DESI J000511.53$-$075558.6	&1.29805	&$-$7.93294 &	0.35\tablenotemark{p}	&104.93	&57.3	&0.79	&25.64	&138	&37.4	&34	&174	&$-$23.88	&8.57  & 1.69 &FR-II,WAT\\
J000533.71$-$032416.7	&DESI J000533.71$-$032416.5	&1.39048	&$-$3.40459 &	0.68\tablenotemark{p}	&43.98	&17.37	&1.22	&26.03	&163.6	&40.5	&30	&218	& --&--	& 0.35	&FR-I,WAT\\
\enddata
\tablecomments{Column (1): source name (JHHMMSS.ss+DDMMSS.s) in the FIRST image provided by RGCMT model, flags a, b, c, d, e, f, g, h, and j denote the sources present in \citet{2011AJ....141...88W}, \citet{2011ApJS..194...31P}, \citet{2017ApJ...844...78P}, \citet{2019AA...626A...8M}, \citet{2021ApJS..254...30P}, \citet{2022ApJS..259...31S}, \citet{2017ApJS..230...20A}, \citet{2017MNRAS.466.4346M}, and \citet{2023DIB....4708974G}, respectively. Column (2): name of the host galaxy if available (JHHMMSS.ss+DDMMSS.s). Column (3): R.A. (J2000, deg). Column (4): Decl. (J2000, deg). Column (5): redshift, \tablenotemark{p} represents photometric redshift, \tablenotemark{s} represents spectroscopic redshift. Column (6): total flux density at 1.4 GHz in mJy. Column (7): total flux density at 3 GHz in mJy. Column (8): spectral index between 1.4GHz and 3 GHz. Column (9): logarithmic luminosity at 1.4 GHz in W ${\rm Hz}^{-1}$. Column (10): opening angle. Column (11): radius of curvature of the radio jets. Column (12): largest angular size. Column (13): largest linear size. Column (14): absolute r-band magnitude. Column (15): logarithmic mass of the black hole. Column (16): $g$-$r$ color values in mag. Column (17): morphological type of the source (FR-I/-II,WAT/NAT). The full table is available in the online journal and via\dataset[DOI: 10.5281/zenodo.14271760]{https://doi.org/10.5281/zenodo.14271760}..
}
\end{deluxetable}
\end{rotatetable*}


\subsection{Spectral Index} \label{sec:spec_index}
\begin{figure}[!ht]
\centering
\plotone{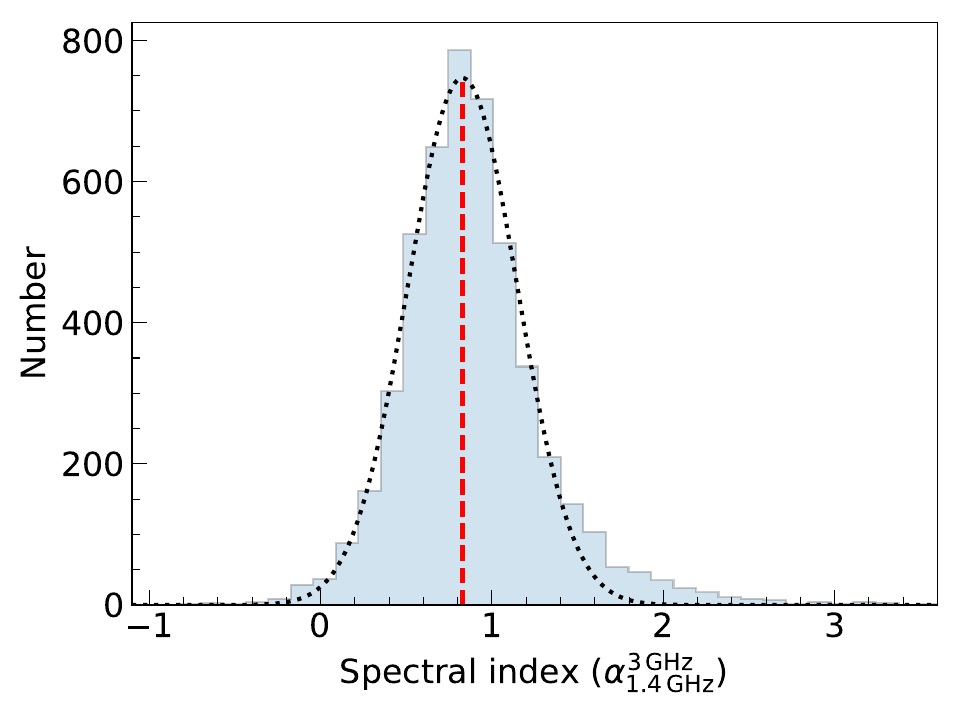}
\caption{A histogram depicting the distribution of the spectral index ($\alpha_{1.4\,{\rm GHz}}^{3\,\rm {GHz}}$) for all BTRGs is presented. The black dotted curve represents a Gaussian fit to the histogram, characterized by a mean of 0.83$\pm$0.01 and a standard deviation of 0.32. The red dashed line indicates the mean spectral index value.
\label{fig:spix_ht}}
\end{figure}

The two-point spectral indexes ($\alpha_{1.4\,{\rm GHz}}^{3\,\rm {GHz}}$) of the BTRGs are calculated using the formula $S_\nu \propto \nu^{-\alpha}$, where $S_\nu$ represents the radiative flux density at a specified frequency, $\nu$, and $\alpha$ denotes the spectral index.
The final spectral index in the BTRGcat is provided for 4852 entries, with 24 BTRGs remaining undetected at 3 GHz within the VLASS data.  
The distribution of the spectral index ($\alpha_{1.4\,{\rm GHz}}^{3\,\rm {GHz}}$) values for all BTRGs is shown in Figure \ref{fig:spix_ht}. The average and median spectral indexes of the full BTRGs are 0.83$\pm$0.01 and 0.85, respectively, with a standard deviation of 0.32. This is steeper than those reported in previous research, which typically find a spectral index of around 0.71 across these frequencies \citep{2012A&A...544A..38Z,2021ApJS..255...30G,2021MNRAS.507.2643A}. However, the range can extend from 0.67 \citep{2017A&A...602A..54M} to 0.85 \citep{2016MNRAS.462.2934V}. We note that due to the lower sensitivity of VLASS to extended emission the spectral indices should be taken as upper limits, i.e. the radio spectra may actually be flatter.

Based on the spectral indices measured between 1.4 GHz and 3 GHz, BTRGs can be categorized into three distinct groups: flat-spectrum sources, which have an index $\alpha_{1.4\,{\rm GHz}}^{3\,\rm {GHz}}<0.5$ and constitute 14.8\% of the sample; steep-spectrum sources, characterized by an index within the range $0.5<\alpha_{1.4\,{\rm GHz}}^{3\,\rm {GHz}}<1.2$ and representing 67.0\% of the sample; and ultra-steep spectrum sources, with indexes $\alpha_{1.4\,{\rm GHz}}^{3\,\rm {GHz}}>1.2$, accounting for 18.2\% of the sample. This indicates that the source population of BTRGs is very diverse in their source properties.

\subsection{Radio Luminosity} \label{sec:Lrad}
\begin{figure*}[ht!]
\centering
\includegraphics[scale=0.5]{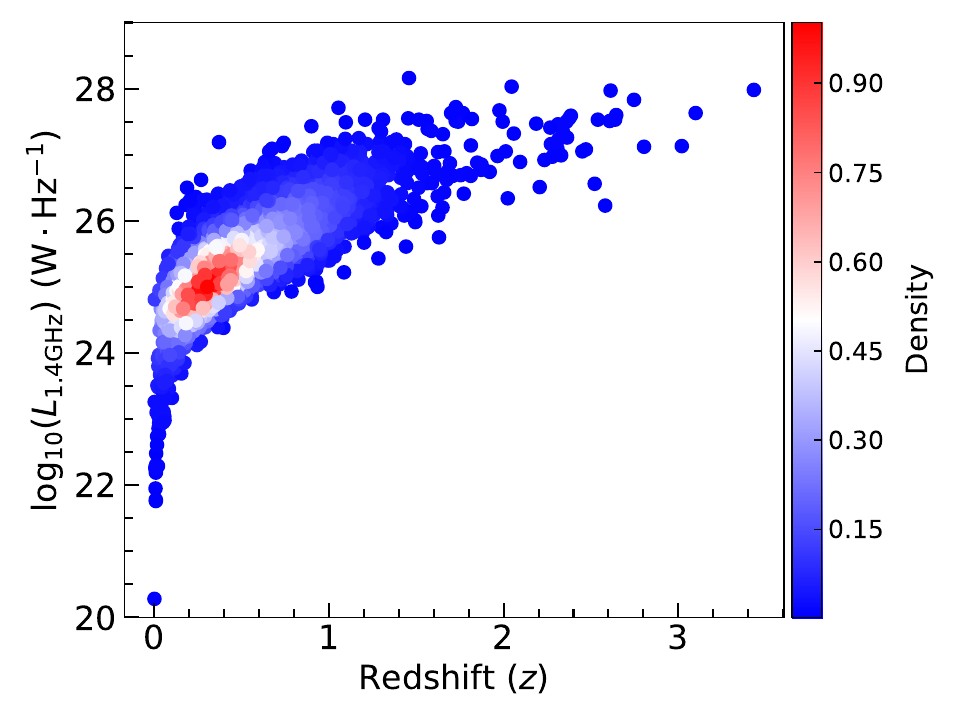}
\includegraphics[scale=0.5]{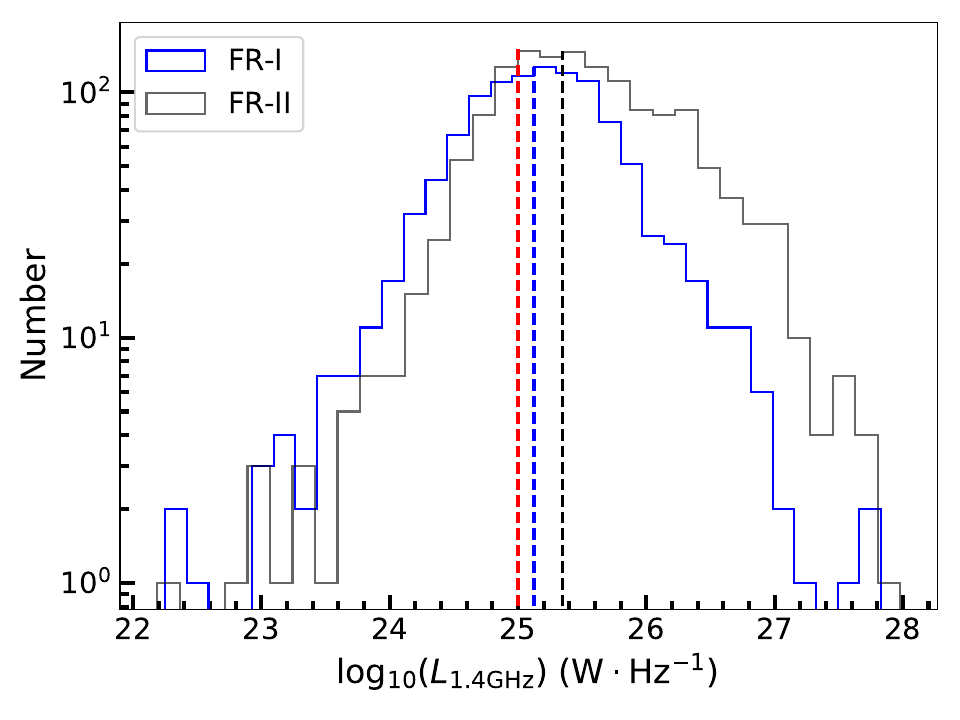}
\caption{Left: the distribution of logarithmic 1.4 GHz luminosity (${\rm log_{10}}(L_{1.4{\rm GHz}})$) for all BTRGs with redshift ($z$) is depicted, where the color of each circle corresponds to the BTRG number, reflecting their respective redshift and 1.4 GHz luminosity values. Right: the distribution of logarithmic 1.4 GHz luminosity (${\rm log_{10}}(L_{1.4{\rm GHz}})$) for FR-I and FR-II sources in our BTRG catalog. The red dashed line represents the classical luminosity break value ${\rm log_{10}}(L_{1.4{\rm GHz}}/{\rm W}\,{\rm Hz}^{-1})=25$. The blue and black dashed lines denote luminosities at peaks of the distribution for FR-I and FR-II, respectively.
\label{fig:Lrad_ht_fr1_fr2}}
\end{figure*}

We quantified the 1.4 GHz luminosity for all BTRGs, utilizing the standard formula \citep[e.g.][]{2009MNRAS.392..617D}, when both the spectral index ($\alpha_{1.4\,{\rm GHz}}^{3\,\rm {GHz}}$) and redshift ($z$) were available. Figure \ref{fig:Lrad_ht_fr1_fr2} left panel shows the distribution of logarithmic 1.4 GHz luminosity (${\rm log_{10}}(L_{1.4{\rm GHz}})$) for full BTRGs. The range of ${\rm log_{10}}(L_{1.4\,{\rm GHz}}/{\rm W\,Hz^{-1}})$ in the BTRG sample is 20.28 to 28.16. The mean and median values of 1.4 GHz luminosity for our BTRGs are ${\rm log_{10}}(L_{1.4\,{\rm GHz}}/{\rm W\,Hz^{-1}})=$25.38 and ${\rm log_{10}}(L_{1.4\,{\rm GHz}}/{\rm W\,Hz^{-1}})=$25.33, respectively. The mean and median values of ${\rm log_{10}}(L/{\rm W\,Hz^{-1}})$ for BTRGs, as reported in \citet{2019AA...626A...8M}, are 25.40 and 25.35, respectively. In \citet{2022MNRAS.516..372B}, these values are 25.72 and 25.73, and in \citet{2021arXiv210315153P}, they are 25.59 and 25.60. These findings suggest that our BTRGs exhibit comparable luminosities to those found in the FIRST, TGSS, and LoTSS catalogs. In our BTRG catalog, 1107 out of 1221 FR-Is and 1421 out of 1799 FR-IIs have been successfully computed for $L_{1.4{\rm GHz}}$. Figure \ref{fig:Lrad_ht_fr1_fr2} right panel presents the distribution of ${\rm log_{10}}(L_{1.4{\rm GHz}})$ for these FR-Is and FR-IIs. Among them, there are 677 FR-I sources with ${\rm log_{10}}(L_{1.4\,\rm GHz}/\rm W\,\rm Hz^{-1}) > 25$, and 330 FR-II sources with ${\rm log_{10}}(L_{1.4\,\rm GHz}/\rm W\,\rm Hz^{-1}) < 25$. This implies that the radio luminosity break between FR classes (${\rm log_{10}}(L_{1.4{\rm GHz}}/{\rm W}\,{\rm Hz}^{-1}) = 25$) does not match their visual classifications as FR-I or FR-II. This finding is consistent with recent results \citep[e.g.][]{2017A&A...601A..81C,2019MNRAS.488.2701M} suggesting that the classical luminosity break by \citet{1974MNRAS.167P..31F} is not necessarily a true FR class divider. This result also reinforces the idea that the luminosity break is not always valid, especially in cases where bent morphology adds ambiguity to classification efforts. Moreover, the distribution of the two classes in the right panel of Figure \ref{fig:Lrad_ht_fr1_fr2} shows that the FR-I distribution reaches its peak at lower luminosities compared with the FR-II distribution and declines much more steeply at high luminosity. This indicates that some evidence of the `standard' model regarding the luminosity of the two classes still exists. Finally, it should be noted that uncertainties in identifying the appropriate hosts, which affect the redshifts of the candidates, as well as inaccuracies in their redshift measurements (particularly for photometric redshifts), also contribute to the variation in the luminosity distributions of the two FR classes. 


\subsection{Opening Angle and Radius of Curvature of the Radio Jets} 
\label{sec:oa_rpa}
\begin{figure*}[ht!]
\centering
\plotone{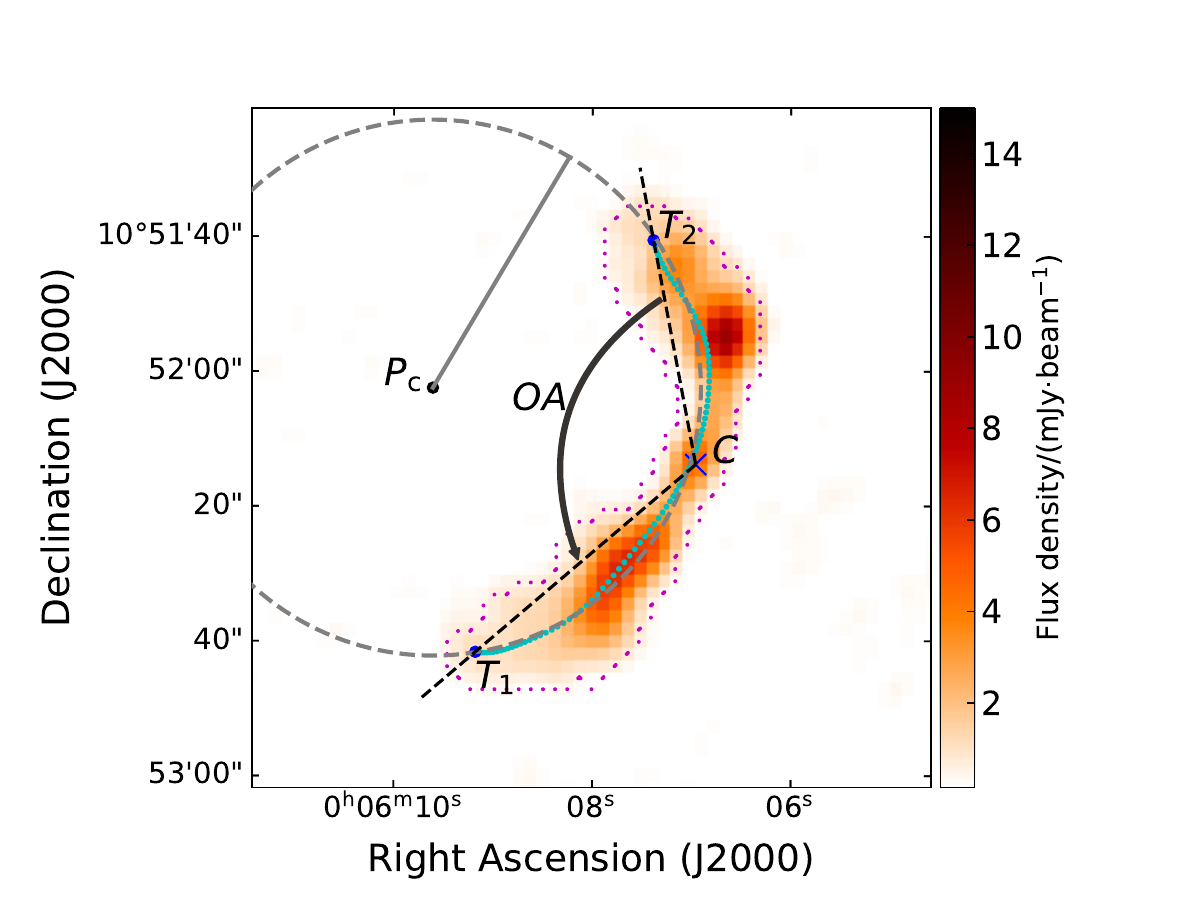}
\caption{Definition example of opening angle ($OA$) for a BTRG. The magenta dotted line represents the predicted polygon points of the BTRG, derived from the transformation of its prediction mask. The green dotted line indicates the center line of the predicted polygon points. The core position ($C$) of the BTRG is denoted by a blue `$\times$’. The tail end positions ($T_1$ and $T_2$) are indicated by blue filled circles. The gray dashed line illustrates the circle of curvature of the radio jets. The black point $P_{\rm c}$ is the center of the fitted circle, and the gray solid line represents the radius of curvature.
\label{fig:OA_exam}}
\end{figure*}

As shown in Figure \ref{fig:OA_exam}, the measurement of the $OA$ for our BTRGs was conducted in three stages. Initially, the predicted mask of each BTRG was transformed into polygon points. Subsequently, a Voronoi diagram \citep{aurenhammer1991voronoi} was generated for the polygon points of each BTRG, which was then employed to determine a center line after applying a Gaussian smoothing filter. The initial and terminal points were designated as the two tail end positions ($T_1$ and $T_2$) of the BTRG. We opted to utilize either the host position or the center position of the center-line as the core position ($C$). In the event of a host galaxy, we selected the host position; conversely, if no host galaxy is present, we selected the center position of the center-line. Finally, the $OA$ was calculated using Equation \ref{eq:oa} as follows:    
\begin{equation}\label{eq:oa}
OA = \arccos \left( \frac{{L^2_{CT_1}}+{L^2_{CT_2}}-{L^2_{T_1T_2}}}{2{L_{CT_1}}\cdot{L_{CT_2}}} \right),
\end{equation}
where $L_{CT_1}$ represents the length of the straight line connecting points $C$ and $T_1$, $L_{CT_2}$ denotes the length of the line segment joining points $C$ and $T_2$, $L_{T_1T_2}$ is the length of the straight line that connects points $T_2$ and $T_1$. The term `arccos' denotes the inverse cosine function. Through the calculation based on Equation \ref{eq:oa}, the $OA$ of the BTRG in Figure \ref{fig:OA_exam} is obtained, which is approximately 120$^\circ$. 
  
\begin{figure*}[ht!]
\centering
\plotone{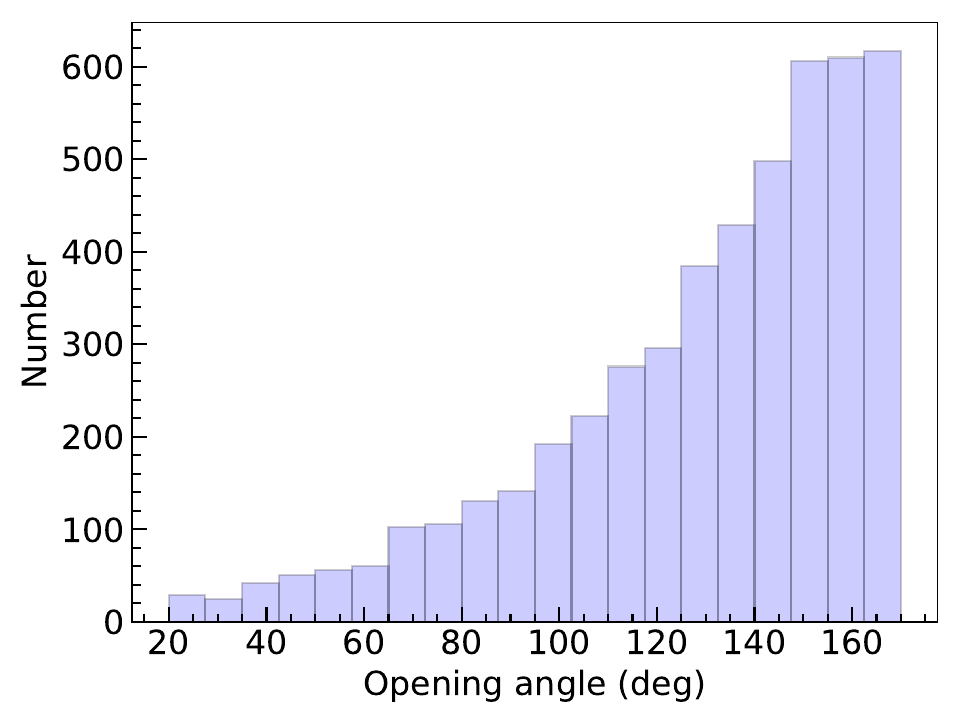}
\caption{The distribution of $OA$ for all BTRG candidates.
\label{fig:OA_ht}}
\end{figure*}

Figure \ref{fig:OA_ht} illustrates the distribution of the $OA$ for all BTRG candidates. We selected as BTRGs only sources with an $OA$ less than 170$^\circ$. The mean and median values of $OA$ are 129.6$^\circ$ and 138.2$^\circ$, respectively. Notably, 652 of these BTRGs have $OA$ values less than 90$^\circ$ and are thus classified as NATs, while 4224 BTRGs have $OA$ values greater than 90$^\circ$ and are thus classified as WATs.  

For the $R_c$ of each BTRG sample, we initially fit an optimal circle through the data points of the center line using the linear least squares fitting method. This fitting process yielded the center point of the fitted circle, see the point $C$ in Figure \ref{fig:OA_exam}. We then calculate the algebraic distance between the center-line data points and the fitted circle centered at point $C$, and the mean of these distances as $R_c$. 
As shown in Figure \ref{fig:OA_exam}, the gray dashed line and solid line respectively indicate the optimal circle and the radius of this circle fitted to the radio jets of the BTRG. The radius of curvature of the radio jets for this BTRG is $R_c$=40$''$.

\subsection{Host Properties}
\begin{figure*}[ht!]
\centering
\includegraphics[scale=0.5]{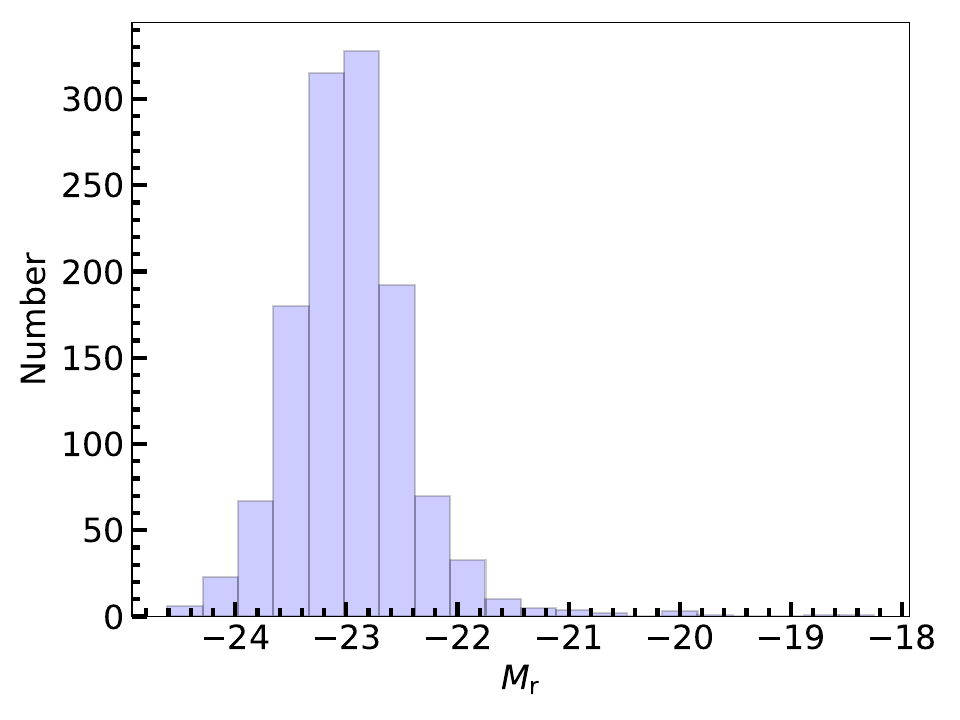}
\includegraphics[scale=0.5]{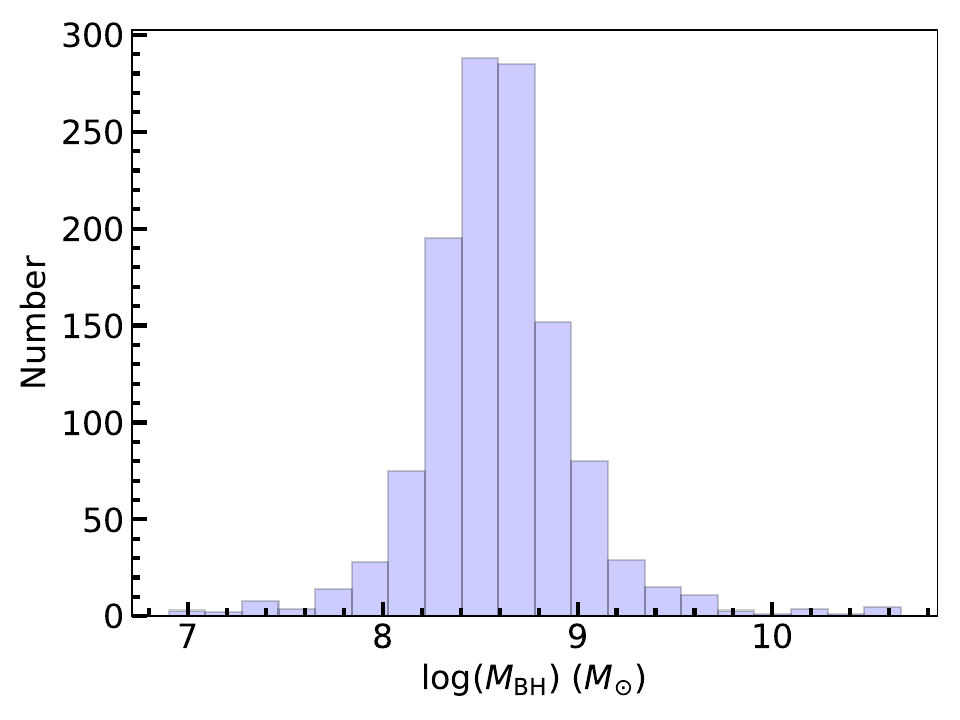}
\caption{Distributions of the absolute r-band magnitude ($M_{\rm r}$) (left panel), and mass of the black hole ($M_{\rm BH}$) (right panel).
\label{fig:Mr_MBH}}
\end{figure*}

We retrieved the values of $M_{\rm r}$ and derived the values of $M_{\rm BH}$ for our BTRGs from the spectral data in SDSS DR16. The $M_{\rm r}$ values of 1241 BTRGs were obtained successfully, while for the remaining ones, they were not obtained because there were no $M_{\rm r}$ values in the SDSS DR16 database. The distribution of the $M_{\rm r}$ of these BTRGs is shown in the left panel of Figure~\ref{fig:Mr_MBH}, given that the hosts cover a range of $-18.25 \gtrsim M_{\rm r} \gtrsim -24.61$. Among them, the mean and median of $M_{\rm r}$ are -22.96 and -23.01 respectively. Almost all of them (99.3\%) have $-21 \gtrsim M_{\rm r} \gtrsim -25$. The $M_{\rm BH}$ were successfully computed for 1203 BTRGs. For the remaining BTRGs, due to the absence of their stellar velocity dispersion values in the SDSS DR16 database, their $M_{\rm BH}$ values were not estimated. The distribution of $M_{\rm BH}$ for these BTRGs is displayed in the right panel of Figure~\ref{fig:Mr_MBH}. The mean and median of ${\rm log_{10}}(M_{\rm BH})$ are 8.60 and 8.58 respectively. Almost all of them (97.8\%) have $9.8 \gtrsim {\rm log_{10}}(M_{\rm BH}) \gtrsim 7.5$ $M_{\odot}$.

\begin{figure*}[htbp] 
\plotone{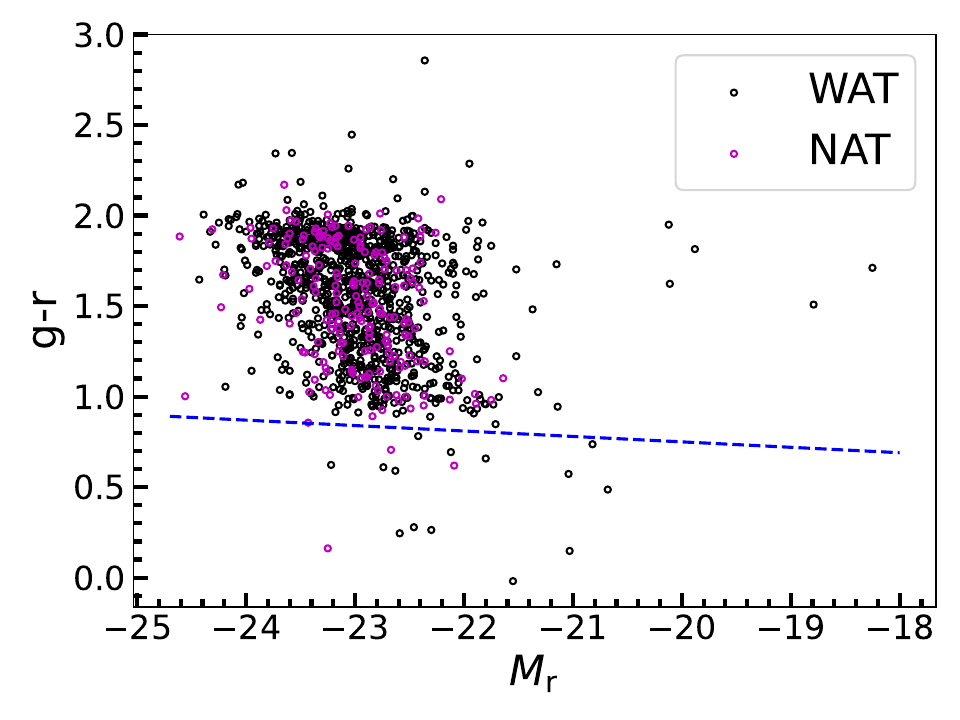}
\caption{$g-r$ color versus absolute $r$-band magnitude ($M_r$) for the BTRGs hosts. The blue dashed line represents the relation from \citet{2006MNRAS.366....2W} that separates the galaxies into red and blue ones.
\label{fig:gr_Mr}}
\end{figure*}

Figure \ref{fig:gr_Mr} presents the distribution of the $g$-$r$ color with respect to the $M_r$ for the hosts of BTRGs. The $g$-$r$ color values were measured from DESI LS DR10. The $g$-$r$ values were obtained for 1201 hosts. The remaining BTRGs did not have their $g$-$r$ color values retrieved because of a lack of relevant DESI LS DR10 data for them. The blue dashed line in Figure \ref{fig:gr_Mr} is utilized to classify the host galaxies into red and blue galaxies. This classification method is from \citet{2006MNRAS.366....2W}. The hosts above the line are classified as red galaxies, while those below the line are classified as blue galaxies. Most of the hosts (98.6\%) appear to have high $g$-$r$ values and are above the blue dashed line; they are, therefore, red galaxies. This is in agreement with the results in \citet{2019AA...626A...8M}, which showed that the BTRGs are typically associated with red galaxies. Based on the same classification scheme, the remaining 17 BTRG hosts (14 WATs and 3 NATs) are classified as blue galaxies. We have not found any common properties among these blue galaxies in the information of our catalog in Table \ref{table:TRGcat}.

\subsection{Association of BTRGs with Clusters}

BTRGs have been demonstrated to be associated with massive clusters \citep{2015MNRAS.446.3310M}. Galaxy clusters, the largest known gravitationally bound structures in the Universe, comprise hundreds to thousands of galaxies held together by gravity \citep[e.g.][]{2012ARA&A..50..353K}. Dark matter makes up about 80\% of the total mass in galaxy clusters and plays a significant role in their formation \citep{2004MNRAS.351..237R}. As a quintessential tool for cosmologists, galaxy clusters facilitate the exploration of the Universe's composition and evolutionary history on a grand scale \citep{2011ARA&A..49..409A,2012MNRAS.424..232W}. A deep understanding of galaxy clusters is instrumental in mapping the large-scale structure of the cosmos \citep{2019ApJ...879....9S,2020AA...633A..19B,2024A&A...686A.196S}.

To this end, we initiate the process by cross-matching our BTRGs that have hosts with the galaxy cluster catalog as presented in \citet{2024ApJS..272...39W}, aiming to identify clusters associated with each BTRG within a search radius of 3$'$. We then select the nearest match within the search radius between these two catalogs. For the remaining unmatched BTRGs, we proceed to utilize NED to search for galaxy clusters associated with each BTRG, also within a search radius of 3$'$. From the search outcomes, we select the clusters that are nearest to the BTRGs. 

Finally, a total of 3286 BTRGs have been identified as corresponding to known galaxy clusters, as listed in Table \ref{table:HT_Cl}. Columns 1 and 2 list the radio names of the BTRGs and the names of their associated clusters. Column 3 gives the redshift values of the associated clusters. Columns 4 and 5 list the comoving distance $D_{\rm c}$ (in Mpc) at the cluster's redshift and the angular separation (in arcsec) between the positions of BTRG hosts and the centres of their associated clusters. Column 6 represents the linear distance $D_{l}$ (in kpc) of the BTRG host from the cluster centre. The radius $r_{500}$ (in Mpc) of the associated clusters is given in column 7. This radius represents the radius enclosing a volume with a mean density that is 500 times the critical density at that redshift. Columns 8 and 9 provide the richness and mass of clusters. Column 10 presents the number of member galaxy candidates within $r_{500}$. A total of 3100 associated clusters were retrieved from the catalog presented in \citet{2024ApJS..272...39W}, of which 582 do not list any previous detection of their cluster by other catalogs and were marked with a `WH'. Additionally, 151 associated clusters were identified from NED. The associated clusters have been identified in 41 known cluster catalogs, with their details summarized in Table \ref{table:HT_Cl_num}.
A total of 3087 BTRGs are found within 1 Mpc of associated clusters. Among these, 1793 BTRGs are situated within a linear distance of 30 kpc, corresponding to the typical size of elliptical galaxies \citep{2010MNRAS.409.1362D}. Importantly, we discovered that a total of 1825 are within their associated cluster, with $D_l$ being less than $r_{500}$, and the redshift differences between the hosts of BTRG and their associated clusters are less than 0.01. Among them, the positions of 1619 BTRG hosts are coincident with the centres of their associated galaxy clusters, with their separations being less than 1$''$. Furthermore, for 94 BTRG hosts, it remains undetermined whether they are within the associated galaxy clusters because there is no $r_{500}$ information for the associated galaxy clusters. This indicates that below half (43.5\%) of our BTRGs that have hosts belong to their associated galaxy clusters. Most BTRGs are not associated with known clusters, which may be due to the selection effects of our BTRGs \citep{2018MNRAS.481.5247O}.

\begin{deluxetable*}{lccccccccc}[ht!]
\tabletypesize{\scriptsize} %
\tablecaption{Cluster details for BTRGs \label{table:HT_Cl}}
\tablehead{
\colhead{Name}  &  \colhead{Cluster} & \colhead{$z_{\rm cl}$}  & \colhead{$D_{\rm c}$} & \colhead{Sep} & \colhead{$D_{l}$} & \colhead{$r_{500}$} & \colhead{$R_{L*,500}$} & \colhead{$M_{500}$} & \colhead{$N_{500}$} \\
\colhead{} &\colhead{} & \colhead{}  & \colhead{(Mpc)} & \colhead{(arcsec)} & \colhead{(kpc)} & \colhead{(Mpc)} & \colhead{} & \colhead{($\times10^{14}M_{\odot}$)} & \colhead{}  
\\
\colhead{(1)} &\colhead{(2)} & \colhead{(3)}  & \colhead{(4)} & \colhead{(5)} & \colhead{(6)} & \colhead{(7)} & \colhead{(8)} & \colhead{(9)} & \colhead{(10)}
}
\startdata
J000115.37$-$082639.6	&AMF J000115.1$-$082646	&0.3287\tablenotemark{s}	&1344	&0	&0	&0.639	&23.35	&1.06	&13 \\
J000121.53$+$010147.4	&WHL J000121.5$+$010149	&0.552\tablenotemark{s}	&2124	&0	&0	&0.662	&32.52	&1.47	&21 \\
J000122.13$-$001135.4	&WHL J000121.9$-$001138	&0.4641\tablenotemark{s}	&1829 &	7	&42	&0.47	&7.22	&0.36	&4 \\
J000331.85$+$002812.0	&maxBCG J000330.7$+$002757	&0.1951\tablenotemark{s}	&827	&0	&0	&0.791	&40.09	&1.8	&17 \\
J000353.42$+$121031.1	&WH J000344.8$+$120949	&0.7438\tablenotemark{p}	&2713 &	129	&973	&0.436	&12.34	&0.57	&6 \\
J000443.77$+$062411.4	&WHL J000441.9$+$062135	&0.5725\tablenotemark{p}	&2190 &	152	&1026	&0.895	&74.21	&3.27	&29 \\
J000511.32$-$075556.9 &WHL J000511.5$-$075559 &0.33\tablenotemark{s} &1349 &0 &0 &0.826 &50.31 &2.24 &16 \\
J000533.71$-$032416.7 &WH22 J000527.1$-$032354 &0.6562\tablenotemark{p} &2453 &102 &732 &0.501 &19.47 &0.89 &12 \\
J000605.20$+$011009.5 &WH J000604.9$+$011001 &1.0397\tablenotemark{p} &3499 &0 &0 &0.42 &16.1 &0.74 &8 \\
J000607.55$-$103028.1	&WHL J000615.3$-$103043	&0.5476\tablenotemark{s}	&2109 &	115	&760	&0.957	&89.53	&3.92	&56 \\
\enddata
\tablecomments{Column (1): source name (JHHMMSS.ss+DDMMSS.s) of the BTRGs. Column (2): cluster name with J2000 coordinates. Column (3): cluster redshift, \tablenotemark{p} represents photometric redshift, \tablenotemark{s} represents spectroscopic redshift. Column (4): comoving distance. Column (5): angular separation between BTRGs and their associated cluster centres. Column (6): linear distances between the positions of BTRG hosts and the centres of their associated clusters. Column (7): cluster radius. Column (8): cluster richness. Column (9): cluster mass. Column (10): number of member galaxy candidates within $r_{500}$. The complete table is available in the online journal and via\dataset[DOI: 10.5281/zenodo.14271760]{https://doi.org/10.5281/zenodo.14271760}.
}
\end{deluxetable*}

The bending of the jet is influenced by the mass of the parent cluster through two mechanisms. Firstly, the density of the ICM is directly proportional to the cluster mass, with more massive clusters possessing a denser ICM. Secondly, the velocity dispersion of BTRGs is likewise correlated with the cluster mass. It is postulated that BTRGs display accelerated motion within high-mass clusters \citep{2010MNRAS.406.2578M, 2015MNRAS.446.3310M}.
The cluster mass $M_{500}$ (in units of $10^{14}M_{\odot}$) for the majority of associated clusters (3100 out of a total of 3286) were derived from the catalog by \citet{2024ApJS..272...39W}. In contrast, the $M_{500}$ for 186 associated clusters was estimated using the cluster richness ($R_{L*,500}$), a parameter that exhibits a strong correlation, as suggested by \citet{2015ApJ...807..178W}:
\begin{equation}\label{eq:m500}
{\rm log}_{10}(M_{500}) = 1.08{\rm log}_{10}(R_{L*,500})-1.37.
\end{equation}

\begin{figure*}[ht!] 
\plotone{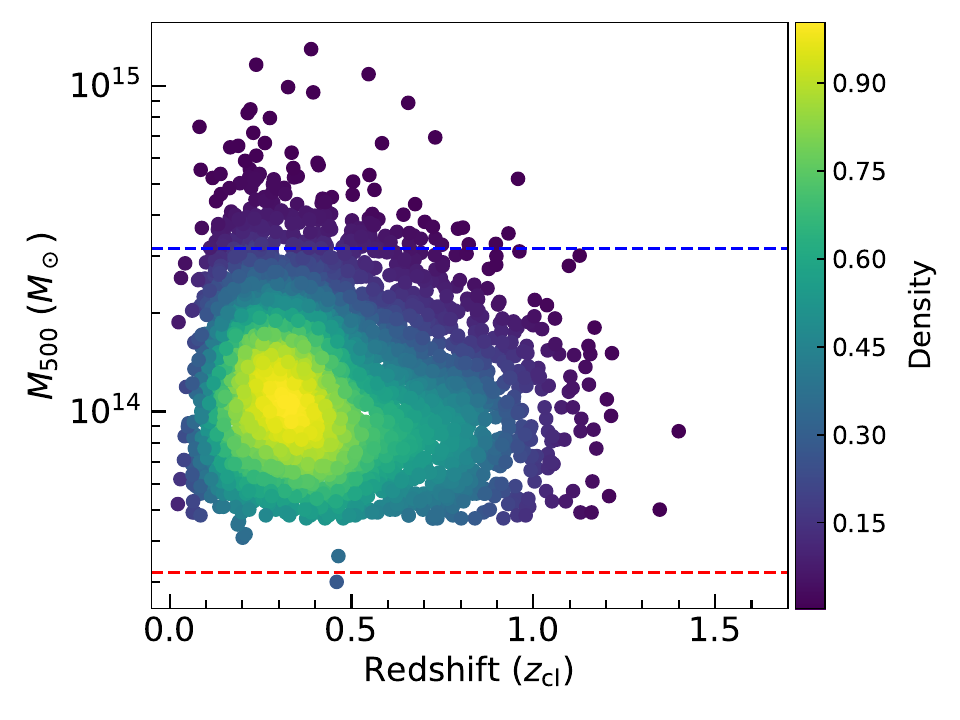}
\caption{Mass vs. redshift distribution of associated clusters. The red and blue lines correspond to the values of $M_{500}=0.32\times10^{14}M_\odot$ and $M_{500}=10^{14.5}M_\odot$, respectively. Only the values of $M_{500}$ provided in Table \ref{table:HT_Cl} are shown.
\label{fig:masses_z}}
\end{figure*}

A comparison of redshifts and masses of associated clusters is shown in Figure \ref{fig:masses_z}. The optical mass ($M_{500}$) of associated clusters spans from $0.30\times10^{14}M_\odot$ to $12.95\times10^{14}M_\odot$, with only one cluster having masses less than $0.32\times10^{14}M_\odot$. This finding aligns with the conclusion drawn by \citet{2015MNRAS.446.3310M} that no BTRGs are found in clusters with masses less than $10^{13}M_\odot$ and only a few in clusters with masses below $10^{13.5}M_\odot$. The mean and median values of $M_{500}$ for the associated clusters are $1.54\times10^{14}M_\odot$ and $1.27\times10^{14}M_\odot$, respectively. Only 198 (6\%) of BTRGs reside in clusters with a mass $M_{500}$ exceeding $10^{14.5}M_\odot$, suggesting that our BTRGs are predominantly located in host clusters of lower mass than anticipated by \citet{2015MNRAS.446.3310M}. However, this finding is in agreement with the measurements reported by \citet{2018MNRAS.481.5247O}.

The velocity dispersion of galaxies within a cluster is proportional to the square root of the cluster mass through the gas temperature \citep{2015MNRAS.446.3310M,2019AJ....157..126G}, hence $M_{500}$ acts as a proxy for the relative velocity between the source and the ICM. In our sample of associated galaxy clusters, we determined that the mean and median values of the logarithmic ratio between ${\rm log}_{10}(M_{500})$ and ${\rm log}_{10}(r_{500})$ both equal 2.43$\times10^{14}M_{\odot}/{\rm Mpc}$. Consequently, we posit that the ICM density remains constant to a first-order approximation with respect to cluster mass and does not have a significant influence on our BTRGs. This conclusion agrees with the findings regarding BTRGs presented in \citet{2019AJ....157..126G}.

\section{Conclusions} \label{sec:cons}
In this paper, we searched for BTRGs with a method that combines a deep learning-based source finder (RGCMT) with visual inspection and constructed a catalog of 4876 BTRGs from the FIRST survey, 3871 of which were newly discovered. These newly discovered BTRGs will significantly increase the number of sources of this type and thus help expand the understanding of such sources.
We found the optical counterparts of 4193 BTRGs by using the combined cross-match and visual inspection methods. The BTRGs in the present catalog have radio luminosities in the range of $1.91\times10^{20} \leq L_{\rm 1.4\,GHz} \leq 1.45\times10^{28}$ ${\rm W\,Hz^{-1}}$, and redshifts up to 3.43. Based on the $OA$ of the two jets, we divided our BTRGs into the WAT and the NAT. Among them, 652 were NATs and 4424 were WATs. According to the radio morphology, we classified our BTRGs into two types: FR-I and FR-II. Among all
BTRGs, 1221 sources were of FR-I type and 1799 sources were of FR-II type. 
A steep spectral index between 1.4 and 3 GHz was found for 67\% of the BTRGs.

The hosts of BTRGs were found to have $M_r$ concentrated in the range from -25 to -21 mag, and $M_{\rm BH}$ spanning from $10^{7.5}$ to $10^{9.5}$ $M_\odot$. Most of the hosts (98.6\%) were red galaxies. By cross-matching with the existing galaxy cluster catalogs, a total of 3286 hosts of BTRGs were found to be associated with galaxy clusters within 3$'$. Among them, 1825 BTRGs (43.5\%) resided in galaxy clusters. All of the associated clusters had a mass lower than the theoretical values. These findings will be of great help in the studies of the environment of BTRGs.

The BTRG catalog presented in this paper is obtained from all FIRST data through deep learning acceleration and meticulous visual inspection, representing the largest and most comprehensive BTRG catalog to date. This catalog will be helpful for future statistical and scientific studies.
We also found that the results provided by the RGCMT model enable key physical properties, such as total flux density, $OA$ and $R_c$, to be calculated automatically. This automated approach is expected to play an important role in the statistical analysis of a large number of radio galaxies.

\section*{Acknowledgments}

The authors thank the anonymous referee for providing a
helpful report. This work was supported by the National SKA Program of China (2022SKA0120101, 2022SKA0130100, 2022SKA0130104), the National Natural Science Foundation of China (Nos. 12103013), the Foundation of Science and Technology of Guizhou Province (Nos. (2021)023), the Foundation of Guizhou Provincial Education Department (Nos. KY(2021)303, KY(2020)003, KY(2023)059). X.Y. was supported by the National Science Foundation of China (12103076, 12233005), the National Key R\&D Program of China (2020YFE0202100), the Shanghai Sailing Program (21YF1455300), and the China Postdoctoral Science Foundation (2021M693267). 

This research work made use of public data from the Karl G. Jansky Very Large Array (VLA); the VLA facility is operated by the National Radio Astronomy Observatory (NRAO). The NRAO is a facility of the National Science Foundation operated under cooperative agreement by Associated Universities, Inc. 

This research has made use of the NASA/IPAC Extragalactic Database (NED), which is funded by the National Aeronautics and Space Administration and operated by the California Institute of Technology.

This publication makes use of data products from the Wide-field Infrared Survey Explorer, which is a joint project of the University of California, Los Angeles, and the Jet Propulsion Laboratory/California Institute of Technology, funded by the National Aeronautics and Space Administration.

This research has made use of the CIRADA cutout service at URL cutouts.cirada.ca, operated by the Canadian Initiative for Radio Astronomy Data Analysis (CIRADA). CIRADA is funded by a grant from the Canada Foundation for Innovation 2017 Innovation Fund (Project 35999), as well as by the Provinces of Ontario, British Columbia, Alberta, Manitoba and Quebec, in collaboration with the National Research Council of Canada, the US National Radio Astronomy Observatory and Australia’s Commonwealth Scientific and Industrial Research Organisation.

This research has made use of the cross-match service \citep{2012ASPC..461..291B,2020ASPC..522..125P} and the VizieR catalogue access tool \citep{2000A&AS..143...23O} provided by CDS, Strasbourg, France. It additionally used the facilities of the Canadian Astronomy Data Centre (CADC), an organization operated by the National
Research Council of Canada with the support of the Canadian
Space Agency. Observations from the VLA, WISE, SDSS,
2dFGRS, 6dFGS, and 2MASS were used in
this work.

Funding for the Sloan Digital Sky Survey IV has been provided by the Alfred P. Sloan Foundation, the U.S. Department of Energy Office of Science, and the Participating Institutions. SDSS-IV acknowledges support and resources from the Center for High Performance Computing  at the University of Utah. The SDSS website is \url{www.sdss4.org}. SDSS-IV is managed by the Astrophysical Research Consortium for the Participating Institutions of the SDSS Collaboration including the Brazilian Participation Group, the Carnegie Institution for Science, Carnegie Mellon University, Center for Astrophysics | Harvard \& Smithsonian, the Chilean Participation Group, the French Participation Group, Instituto de Astrof\'isica de Canarias, The Johns Hopkins University, Kavli Institute for the Physics and Mathematics of the Universe (IPMU) / University of Tokyo, the Korean Participation Group, Lawrence Berkeley National Laboratory, Leibniz Institut f\"ur Astrophysik Potsdam (AIP),  Max-Planck-Institut f\"ur Astronomie (MPIA Heidelberg), Max-Planck-Institut f\"ur Astrophysik (MPA Garching), Max-Planck-Institut f\"ur Extraterrestrische Physik (MPE), National Astronomical Observatories of China, New Mexico State University, New York University, University of Notre Dame, Observat\'ario Nacional / MCTI, The Ohio State University, Pennsylvania State University, Shanghai Astronomical Observatory, United Kingdom Participation Group, Universidad Nacional Aut\'onoma de M\'exico, University of Arizona, University of Colorado Boulder, University of Oxford, University of Portsmouth, University of Utah, University of Virginia, University of Washington, University of Wisconsin, Vanderbilt University, and Yale University.

This research uses services or data provided by the Astro Data Lab, which is part of the Community Science and Data Center (CSDC) Program of NSF NOIRLab. NOIRLab is operated by the Association of Universities for Research in Astronomy (AURA), Inc. under a cooperative agreement with the U.S. National Science Foundation.

This publication makes use of data products from the Two Micron All Sky Survey, which is a joint project of the University of Massachusetts and the Infrared Processing and Analysis Center/California Institute of Technology, funded by the National Aeronautics and Space Administration and the National Science Foundation.

The Legacy Surveys consist of three individual and complementary projects: the Dark Energy Camera Legacy Survey (DECaLS; Proposal ID \#2014B-0404; PIs: David Schlegel and Arjun Dey), the Beijing-Arizona Sky Survey (BASS; NOAO Prop. ID \#2015A-0801; PIs: Zhou Xu and Xiaohui Fan), and the Mayall z-band Legacy Survey (MzLS; Prop. ID \#2016A-0453; PI: Arjun Dey). DECaLS, BASS and MzLS together include data obtained, respectively, at the Blanco telescope, Cerro Tololo Inter-American Observatory, NSF’s NOIRLab; the Bok telescope, Steward Observatory, University of Arizona; and the Mayall telescope, Kitt Peak National Observatory, NOIRLab. Pipeline processing and analyses of the data were supported by NOIRLab and the Lawrence Berkeley National Laboratory (LBNL). The Legacy Surveys project is honored to be permitted to conduct astronomical research on Iolkam Du’ag (Kitt Peak), a mountain with particular significance to the Tohono O’odham Nation.

NOIRLab is operated by the Association of Universities for Research in Astronomy (AURA) under a cooperative agreement with the National Science Foundation. LBNL is managed by the Regents of the University of California under contract to the U.S. Department of Energy.

This project used data obtained with the Dark Energy Camera (DECam), which was constructed by the Dark Energy Survey (DES) collaboration. Funding for the DES Projects has been provided by the U.S. Department of Energy, the U.S. National Science Foundation, the Ministry of Science and Education of Spain, the Science and Technology Facilities Council of the United Kingdom, the Higher Education Funding Council for England, the National Center for Supercomputing Applications at the University of Illinois at Urbana-Champaign, the Kavli Institute of Cosmological Physics at the University of Chicago, Center for Cosmology and Astro-Particle Physics at the Ohio State University, the Mitchell Institute for Fundamental Physics and Astronomy at Texas A\&M University, Financiadora de Estudos e Projetos, Fundacao Carlos Chagas Filho de Amparo, Financiadora de Estudos e Projetos, Fundacao Carlos Chagas Filho de Amparo a Pesquisa do Estado do Rio de Janeiro, Conselho Nacional de Desenvolvimento Cientifico e Tecnologico and the Ministerio da Ciencia, Tecnologia e Inovacao, the Deutsche Forschungsgemeinschaft and the Collaborating Institutions in the Dark Energy Survey. The Collaborating Institutions are Argonne National Laboratory, the University of California at Santa Cruz, the University of Cambridge, Centro de Investigaciones Energeticas, Medioambientales y Tecnologicas-Madrid, the University of Chicago, University College London, the DES-Brazil Consortium, the University of Edinburgh, the Eidgenossische Technische Hochschule (ETH) Zurich, Fermi National Accelerator Laboratory, the University of Illinois at Urbana-Champaign, the Institut de Ciencies de l’Espai (IEEC/CSIC), the Institut de Fisica d’Altes Energies, Lawrence Berkeley National Laboratory, the Ludwig Maximilians Universitat Munchen and the associated Excellence Cluster Universe, the University of Michigan, NSF’s NOIRLab, the University of Nottingham, the Ohio State University, the University of Pennsylvania, the University of Portsmouth, SLAC National Accelerator Laboratory, Stanford University, the University of Sussex, and Texas A\&M University.

BASS is a key project of the Telescope Access Program (TAP), which has been funded by the National Astronomical Observatories of China, the Chinese Academy of Sciences (the Strategic Priority Research Program ``The Emergence of Cosmological Structures’’ Grant \# XDB09000000), and the Special Fund for Astronomy from the Ministry of Finance. The BASS is also supported by the External Cooperation Program of Chinese Academy of Sciences (Grant \# 114A11KYSB20160057), and Chinese National Natural Science Foundation (Grant \# 12120101003, \# 11433005).

The Legacy Survey team makes use of data products from the Near-Earth Object Wide-field Infrared Survey Explorer (NEOWISE), which is a project of the Jet Propulsion Laboratory/California Institute of Technology. NEOWISE is funded by the National Aeronautics and Space Administration.

The Legacy Surveys imaging of the DESI footprint is supported by the Director, Office of Science, Office of High Energy Physics of the U.S. Department of Energy under Contract No. DE-AC02-05CH1123, by the National Energy Research Scientific Computing Center, a DOE Office of Science User Facility under the same contract; and by the U.S. National Science Foundation, Division of Astronomical Sciences under Contract No. AST-0950945 to NOAO.


\vspace{5mm}
\facilities{CDS, VLA, WISE, Sloan, CTIO:2MASS, Astro Data Lab}


\software{The work carried out in this paper made use of the
following software packages and tools: Astropy 
 \citep{2013A&A...558A..33A,2018AJ....156..123A}, APLpy \citep{2012ascl.soft08017R}, Matplotlib \citep{2007CSE.....9...90H}, Astroquery \citep{2019AJ....157...98G}, Numpy \citep{2020Natur.585..357H}, Pandas \citep{mckinney-proc-scipy-2010,reback2020pandas}, scikit-image \citep{van2014scikit}, SciPy \citep{virtanen2020scipy}, SAOImage
DS9 \citep{2003ASPC..295..489J}, TOPCAT \citep{2005ASPC..347...29T}}.

\textit{Note added in proof.} Upon further inspection of the 17 BTRGs with color indices $g-r$
below the blue dashed line in Fig. 10, we found that 12 of them are
spectroscopic QSOs in SDSS, another three are blazars, one is a $z\sim$1
galaxy, and one (J125648.57+481749.8) is a merger of two late-type galaxies
that should be removed from the sample of BTRGs.

\appendix

\section{Details of Associated Galaxy Cluster Catalogs} \label{appx:ht_Cl_num}
The associated clusters of our BTRGs have been identified within 41 previously published cluster catalogs. The details are summarized in Table \ref{table:HT_Cl_num}.
\startlongtable
\begin{deluxetable*}{lccc}
\tabletypesize{\scriptsize}
\tablewidth{0pt}
\tablecaption{Details of various cluster catalogs used in this work  \label{table:HT_Cl_num}}
\tablehead{
\colhead{Catalog Name} & \colhead{Observation Band} &  \colhead{No. Clusters} & \colhead{Reference(s)}  
}
\colnumbers
\startdata
Wen\&Han 2024 (WH) & Optical, SZ, X-ray & 582 &  \citet{2024ApJS..272...39W} \\
Wen\&Han 2022 (WH22) & Optical, SZ, IR, X-ray & 12  & \citet{2022MNRAS.513.3946W} \\
Wen\&Han 2021 (WH21) & Optical, IR & 25  & \citet{2021MNRAS.500.1003W} \\
Wen+Han+Yang (WHY18) & Optical, IR& 58 & \citet{2018MNRAS.475..343W} \\
Wen+Han+Liu (WHL) or Wen\&Han 2015 ([WH2015]) & Optical, X-ray & 954 & \citet{2009ApJS..183..197W, 2012ApJS..199...34W, 2015ApJ...807..178W} \\ 
Max Brightest Cluster Galaxy (MaxBCG) & Optical & 480 &  \citet{2007ApJ...660..239K} \\
Gaussian Mixture BCG (GMBCG) & Optical & 164 & \citet{2010ApJS..191..254H} \\
Fast Search and Find of Density Peaks (CFSFDP) & Optical & 177 & \citet{2021ApJS..253...56Z, 2022RAA....22f5001Z} \\
Cluster finding algorithm based on Multi-band  & Optical & 35 & \citet{2014MNRAS.444..147O} \\ 
Identification of Red sequence gAlaxies (CAMIRA) &  &  &  \\
Adaptive Matched Filter (AMF) & Optical & 277 & \citet{2011ApJ...736...21S} \\
Zwicky Cluster (ZwCl) & Optical & 2 &  \citet{1961cgcg.book.....Z,1963cgcg.book.....Z,1965cgcg.book.....Z,1966cgcg.book.....Z}; \\
 &  &  &  \citet{1968cgcg.bookR....Z} \\
Yang 2021 (Y21) & Optical & 394 & \citet{2021ApJ...909..143Y} \\
Northern Sky optical Cluster Survey (NSCS) or  & Optical & 19 & \citet{2003AJ....125.2064G,2004AJ....128.1017L};\\
 Northern Sky optical Cluster (NSC) &  &  & \citet{2009AJ....137.2981G}\\
SPectroscopic IDentification of  & X-ray & 6 & \citet{2021MNRAS.503.5763K} \\
eROSITA Sources (SPIDERS) &  &  &  \\
XMM X-Ray Cluster Survey (XMMXCS) & X-ray & 1 & \citet{2016ApJ...816...98Z} \\
Wavelet Z Photometric (WaZP) & Optical, IR& 2  & \citet{2021MNRAS.502.4435A} \\
Virgo Cluster (VIRGO) & X-ray & 1 & \citet{2016MNRAS.455..846W,2017ApJ...843...50C} \\
Volume-limited sample 1 Compact Group (V1CG) & IR & 1 & \citet{2017ApJ...835..280L} \\
Sloan Digital Sky Survey  & Optical & 1 & \citet{2005AJ....130..968M} \\
C4 Cluster (SDSS-C4) &  &  &  \\
SDSS-C4 based on Data  & Optical & 1 & \citet{2007MNRAS.379..867V} \\ 
Release 3 (SDSS-C4-DR3) &  &  &  \\ 
Sloan Digital Sky Survey (SDSS) & Optical & 6  & \citet{2015ApJS..219...12A,2021ApJS..253....3H}; \\
 &  &   & \citet{2007ApJ...660.1176E} \\
red-sequence Matched-filter Probabilistic  & Optical, IR & 5 & \citet{2018ApJ...856..172S} \\
 Percolation (RedMapper) &  &  &  \\
RedMapper Cluster (RM) & Optical, IR & 8 & \citet{2015MNRAS.453...38R} \\
RASS-CFHTLS X-ray selected  & X-ray & 1 & \citet{2015ApJ...799...60M} \\
galaxy Clusters (RCC) &  &  &  \\
Planck Sunyaev-Zel'dovich  & SZ & 1 & \citet{2015MNRAS.450..592R}; \\
cluster-2nd release (PSZ2) &  &  & \citet{2016AA...594A..27P} \\
MultiScale Probability Mapping (MSPM) & Optical & 12 & \citet{2012MNRAS.422...25S} \\
Meta-Catalogue of X-ray detected  & X-ray& 1 & \citet{2011AA...534A.109P} \\
Clusters of galaxies (MCXC) & &  &  \\
Kilo Degree Survey DR2 (KDR2) & Optical & 1 & \citet{2017AA...598A.107R} \\
KPNO/Deeprange Distant Cluster  & Optical & 2 & \citet{2002ApJ...579...93P} \\
Survey (KDCS) &  &  &  \\
Gunn+Hoessel+Oke clusters of  & Optical & 1 & \citet{1986ApJ...306...30G} \\
galaxies (GHO) &  &  &  \\
Galaxy Clusters with Masses (GCwM) & X-ray, SZ & 1 & \citet{2016AA...587A.158A} \\
eROSITA Final Equatorial-Depth  & X-ray & 1 & \citet{2022AA...661A...1B} \\
Survey (eFEDS) &  &  &  \\
Clusters from Masses and Photometric  & Optical, IR& 42 & \citet{2024MNRAS.531.2285Y} \\
Redshifts (CLuMPR) & &  &  \\
Chandra Deep Group Survey (CDGS) & X-ray & 1 & \citet{2015MNRAS.447.3723P} \\
Abell Clusters of Galaxies (ABELL) & Optical & 3 & \citet{1958ApJS....3..211A,1989ApJS...70....1A}\\
Estrada+Annis+Diehl ([EAD2007]) & Optical& 1 & \citet{2007ApJ...660.1176E}  \\
Donahue+Scharf+Mack ([DSM2002]) & Optical, X-ray & 1& \citet{2002ApJ...569..689D} \\ 
Canada-France-Hawaii Telescope  & Optical & 3 & \citet{2018AA...613A..67S} \\
 Legacy Survey (CFHTLS) &  &  &  \\
Li+Yee+Hsieh ([LYH2012]) & Optical, IR& 1& \citet{2012ApJ...749..150L}\\
Massive and Distant Clusters of WISE & IR & 1 & \citet{2019ApJS..240...33G} \\
Survey Overdense Object (MOO) &  &  &  \\
ROSAT All-Sky Survey X-ray Galaxy & X-ray & 1 & \citet{2022AA...658A..59X} \\ 
Cluster Catalog (RXGCC) &  &  &  \\ 
\hline
Total & & 3286 & \\
\enddata
\tablecomments{Column (1): the name of the cluster catalog (full name (short name of catalog for Column (2) in Table \ref{table:HT_Cl})). Column (2): the observation band of the cluster catalog (SZ: Sunyaev-Zel'dovich, IR: infrared). Column (3): the number of clusters from each catalog that coincide with one of our BTRGs. Column (4): reference.}
\end{deluxetable*}

\bibliography{sample631}{}
\bibliographystyle{aasjournal}



\end{document}